%% file: PVIS2026-blocking-arXiv.tex
\documentclass[journal,preprint]{vgtc}                    
\onlineid{5494}

\vgtccategory{Research}
\vgtcpapertype{Application/Design Study}

\graphicspath{{./figs/}{./}} 

\usepackage{times}
\usepackage{gensymb}
\usepackage{amssymb}
\usepackage{multirow}
\usepackage{amsmath}
\usepackage{tablefootnote}

\newcommand{\para}[1]{\vspace{1mm}\noindent{\textbf{#1}}}

\newcommand{\etal}{{et al.}}
\newcommand{\eg}{{e.g.}}

\newcommand{\ie}{{i.e.}}

\newcommand{\lat}{\mathrm{latitude}}
\newcommand{\lon}{\mathrm{longitude}}

\usepackage{float}
\title{Spatiotemporal Detection and Uncertainty Visualization of Atmospheric Blocking Events}

\author{
\authororcid{Mingzhe Li}{0000-0003-0355-1919}, 
\authororcid{Peer Nowack}{0000-0003-4588-7832},
and \authororcid{Bei Wang}{0000-0002-9240-0700}
}

\authorfooter{
\item Mingzhe Li is with the University of Notre Dame; this work was completed while at the University of Utah.
  	E-mail: mli33@nd.edu
\item Peer Nowack is with the Karlsruhe Institute of Technology.
      E-mail: peer.nowack@kit.edu
\item Bei Wang is with the University of Utah. 
	E-mail: beiwang@sci.utah.edu
}

\abstract{ 
\input{sec-abstract}
} 


\begin{document}

\maketitle

\input{sec-introduction}

\input{sec-related-work}

\input{sec-background}

\input{sec-data}

\input{sec-method}
\input{sec-implementation}

\input{sec-results}

\input{sec-discussion}

\acknowledgments{
ML and BW were partially supported by U.S. National Science Foundation (NSF) grants IIS-1910733 and IIS-2145499, and by U.S. Department of Energy (DOE) grant DE-SC0021015. PN was supported by the UK Natural Environment Research Council (NERC) grant NE/V012045/1.
The authors thank Dagstuhl Seminar 23342 (Computational Geometry of Earth System Analysis) for initiating this collaboration.}

\bibliographystyle{abbrv-doi}
\bibliography{refs-blocking}

\clearpage
\appendix
\input{sec-runtime}

\input{sec-extended-eval}

\end{document}

%% file: sec-abstract.tex
Atmospheric blocking events are quasi-stationary high-pressure systems that disrupt the typical paths of polar and subtropical air currents, often producing prolonged extreme weather events such as summer heat waves or winter cold spells. Despite their critical role in shaping mid-latitude weather, accurately modeling and analyzing blocking events in long meteorological records remains a significant challenge. 
To address this challenge, we present an uncertainty visualization framework for detecting and characterizing atmospheric blocking events. First, we introduce a geometry-based detection and tracking method, evaluated on both pre-industrial climate model simulations (UKESM) and reanalysis data (ERA5), which represent historical Earth observations assimilated from satellite and station measurements onto regular numerical grids using weather models.
Second, we propose a suite of uncertainty-aware summaries: contour boxplots that capture representative boundaries and their variability, frequency heatmaps that encode occurrences, and 3D temporal stacks that situate these patterns in time. Third, we demonstrate our framework in a case study of the 2003 European heatwave, mapping the spatiotemporal occurrences of blocking events using these summaries. Collectively, these uncertainty visualizations reveal where blocking events are most likely to occur and how their spatial footprints evolve over time. We envision our framework as a valuable tool for climate scientists and meteorologists: by analyzing how blocking frequency, duration, and intensity vary across regions and climate scenarios, it supports both the study of historical blocking events and the assessment of scenario-dependent climate risks associated with changes in extreme weather linked to blocking.

%% file: sec-introduction.tex
\section{Introduction}
\label{sec:introduction}

Atmospheric blocking events are large-scale, persistent weather patterns that occur in the mid-latitudes. Their defining feature is a quasi-stationary high-pressure system that lingers over a geographical region for several days (typically five or more), disrupting the usual paths of the polar jet streams~\cite{Rex1950,Lupo2020}. Blocking events are frequently associated with extreme weather, most notably summer heatwaves and winter cold spells~\cite{Kautz2022,WilkinsonNowackJoshi2025}. By altering the surrounding atmospheric circulation, they are also linked to upstream and downstream extremes, such as the devastating floods in Libya and central Greece in September 2023, which arose from an omega blocking event over Central Europe~\cite{RobertoToniazzo2023}. Analyzing trends in blocking event occurrences, both across continents and in climate change simulations, is therefore of critical importance to the scientific community and society at large. 

However, detecting and characterizing blocking events in long atmospheric time series remains a challenging task. They are characterized by quasi-stationary high-pressure anomalies that remain over geographical regions for typically five or more days. While this may appear straightforward to define and detect at first glance, individual blocking events exhibit highly diverse characteristics~\cite{Kautz2022}, leading to ambiguity in their definition and detection. 
Blocking indices commonly define blocking in terms of different anomalies and gradients of key dynamical variables such as potential vorticity, geopotential height, potential temperature, and sea-level pressure. Consequently, the use of different blocking indices yields substantial discrepancies in both the frequency and spatial distribution of blocking events~\cite{Nowack2021}. 

A further challenge lies in the representation of blocking in state-of-the-art climate models, which are essential for estimating future scenario-dependent changes in climate and atmospheric dynamics. Despite notable advances, including increases in spatial resolution~\cite{Schiemann2017}, climate models still exhibit systematic low biases in blocking frequency compared to observations~\cite{Woollings2018}.

In summary, tracking progress in the modeling and analysis of scenario-dependent future changes in blocking event frequency and characteristics is critical for understanding potential shifts in extreme weather events. This requires robust methods for the automated detection of blocking events in decadal- to centennial-scale climate simulations. Ideally, such methods should also capture detailed event characteristics, including the size of the high-pressure system, its horizontal and vertical structure, its persistence, and its seasonal and regional variability in likelihood of occurrence.

To address these challenges, Thomas et al.~\cite{Nowack2021} recently developed expert-labeled ground truth datasets for atmospheric summer blocking events over Europe, based on approximately 41 years of ERA5 reanalysis observations~\cite{Hersbach2020} and a 101-year climate model simulation~\cite{Sellar2019}. Using this dataset, they demonstrated that a machine learning approach based on self-organizing maps outperforms major blocking indices. However, their method does not provide detailed information on the spatiotemporal characteristics of blocking events, which are essential for long-term analyses of climatological changes in blocking.

Building on the work of Thomas et al.~\cite{Nowack2021}, we aim to advance methodologies for characterizing how blocking frequency, duration, and intensity vary across regions and climate scenarios, thereby enabling a more comprehensive assessment of the risks and spatiotemporal characteristics of prolonged extreme events.
To that end, we propose an uncertainty visualization framework for detecting and characterizing variations in blocking events across multiple spatiotemporal scales.
Our contributions are as follows:
\begin{itemize}[noitemsep,leftmargin=*]
\item \textbf{Geometry-based detection and tracking.} We present a geometry-based pipeline that identifies and tracks high-pressure systems from preprocessed geopotential-height fields, producing results that are both interpretable and reproducible.
\item \textbf{Improved detection quality.} Our framework achieves higher detection accuracy and F1-scores on both reanalysis (ERA5) and climate-model (UKESM) datasets compared to two baseline blocking indices, namely SOM-BI~\cite{Nowack2021} and DG83~\cite{DG83}, when evaluated against expert-labeled ground truth.
\item \textbf{Event-centric representation.} Unlike SOM-BI, which provides pattern labels rather than discrete events, our method explicitly detects and tracks individual blocking events, returning their spatial footprints (boundaries, areas, lifetimes, and trajectories) as the primary representation and backbone for subsequent analysis.
\item \textbf{Uncertainty-aware summaries.} Building on the event representation, we introduce uncertainty-aware visualizations—including contour boxplots, frequency heatmaps, and 3D temporal stacks—that succinctly convey the typical structure, spread, location, and seasonal evolution of blocking events. To the best of our knowledge, these event-level uncertainty summaries represent a novel contribution to the study of blocking and complement existing blocking indices.
\end{itemize}

%% file: sec-related-work.tex
\section{Related Work}
\label{sec:related-work}

\para{Atmospheric blocking indices.}
Atmospheric blocking is strongly associated with extreme weather events in the mid-latitudes. The primary feature of each blocking event is a quasi-stationary high-pressure system, which can be observed in surface pressure variations but typically even more clearly in other atmospheric dynamical variables used to characterize large-scale weather systems, such as mid-tropospheric geopotential height anomalies. Given the nature of high-pressure systems and the associated clear-sky conditions, blocking regionally leads to anomalously warm conditions during summer and cold conditions during brisk winter days. 
In addition, blocking events affect the general background atmospheric flow in their vicinity, leading to anomalous heat transport patterns. As a result, the hottest areas associated with blocking-related heatwaves are often found outside the main high-pressure system. Due to their effect on large-scale atmospheric flow, blocking can also lead to extreme rainfall elsewhere. For example, when an omega blocking event over Central Europe trapped a low-pressure storm over central Greece, it led to days of extreme rainfall and subsequent catastrophic flooding in September 2023~\cite{RobertoToniazzo2023}.

To better understand blocking and to investigate the influence of, \eg, changes in blocking frequencies and duration under climate change, meteorologists have developed a variety of approaches to automatically detect blocking in long atmospheric dynamics time series. 
Since manually labeling events seemed prohibitive, these methods mostly refer to several major blocking indices (BIs) that have been suggested to objectively identify events~\cite{Lejenas1983,DG83,Tibaldi1990,Pelly2003,Nowack2021}. 
Notably, while previous comparisons of BIs show similar climatologies (\ie, maps of long-term average blocking frequencies across a spatial region), studies also found significant discrepancies. These have naturally led to questions concerning how to objectively measure skill in blocking detection, also depending on meteorological season \cite{Croci-Maspoli2007,Barriopedro2010,Pinheiro2019,Nowack2021}. 
Therefore, Thomas et al.~\cite{Nowack2021} introduced a new expert-labeled ground-truth dataset for summer blocking events over Europe. In their comparison of three widely used BIs, the index developed by Dole and Gordon (which we refer to as DG83)~\cite{DG83} performed the best, and they additionally demonstrated that a different approach using machine learning---based on self-organizing maps~\cite{Kohonen2001} to categorize weather states---led to further performance gains in blocking detection. They referred to their index as SOM-BI. In the following, we will use the ground-truth dataset developed in \cite{Nowack2021}, taking the DG83 and SOM-BI methods as our baselines. 

Other frequently used methods to study the climatology and characteristics of blocking include K-means clustering analyses to study weather regimes \cite{Vautard1990,Michelangeli1995,Cassou2008,UllmannFontaine2014,StrommenMavilia2019,Fabiano2021}.
In addition, there have been recent deep-learning efforts for detecting and analyzing blockings, including a two-stage CNN approach that detects and localizes blocking patterns in global climate model output \cite{Muszynski2021} with a BI as ground truth, and a UNet-based model that aims to reconstruct summer blocking frequencies over the Last Millennium from tree-ring-based temperature fields \cite{Karamperidou2024}.
A complementary line of research focuses on forecasting the blocking maintenance using deep learning models. For example, Zhang et al.~\cite{ZhangFinkelAbbot2024} employed transfer learning to train a network that predicts whether a nascent Atlantic block will persist several days after its onset, using the DG83 index as ground truth. This forecasting objective differs from approaches that analyze full temporal sequences to detect and summarize complete blocking episodes.

\para{Tracking atmospheric features.}
Many atmospheric features can be detected and analyzed via feature extraction and tracking. 
Beyond high-pressure systems contributing to blocking events, recent work has tracked cyclones and pressure-perturbations using topology-aware segmentation, with robustness analyses across thresholds and resolutions~\cite{YanGuo2024,YanUllrich2023,WidanagamaachchiJacquesWang2017}. 
Related pipelines track low-level cloud systems by extracting cloud objects/systems via thresholding and assembling trajectories across time~\cite{LiChatterjeeGlassmeier2025, SokolowskyFreemanJones2024,FengHardinBarnes2023}, and track atmospheric rivers by identifying elongated moisture plumes and following them across successive time steps \cite{RalphRutz2019,RutzShields2019}. 
Earlier efforts also track deep-convective clouds directly from satellite images using topological and computer-vision techniques~\cite{DoraiswamyNatarajan2013,Gunnar2003}. 
Across these lines, common ingredients of tracking atmospheric features are region/feature extraction, matching features on consecutive times, treatment of merges/splits, and explicit lifetime/size criteria. 
In this spirit, our blocking detection method adopts a simple region-overlap criterion to enforce quasi-stationarity of high-pressure systems while remaining easy to reproduce and compare.

\para{Uncertainty visualizations for atmospheric events.}
Ensemble visualization methods can summarize how atmospheric events vary across members, times, or models.
Wang \etal~\cite{WangHazarika2019} classified these methods by data types, visual encodings, and analysis tasks, noting that ensembles inherently express uncertainty through multiple realizations of the same quantity;
see other uncertainty-visualization surveys~\cite{PangWittenbrink1997,KamalDhakal2021,PotterRosen2012}.

In meteorology and climatology, uncertainty visualizations have long supported ensemble analysis:
Ensemble-Vis provides interactive overviews and statistical summaries to expose probabilistic characteristics in forecasting and climate applications~\cite{PotterWilson2009}; time-varying weather ensembles have been examined for parameter sensitivity and resolution effects~\cite{BiswasLin2017}; point/curve-based uncertainties have been shown with graduated glyphs and ribbons (Noodles)~\cite{SanyalZhang2010}; and uncertainty-aware visual analysis of atmospheric rivers with topological tools summarizes variability and occurrence across ensembles~\cite{LanGamelin2024}. 
For feature-centric event summaries, where boundaries or tracks are the primary objects, contour boxplots compute order statistics on ensembles of contours to show a representative boundary (median) and central/outer envelopes; the approach extends to streamlines, pathlines, and curve boxplots, and has been used for hurricane tracks and weather-forecast ensembles~\cite{WhitakerMirzargar2013,MirzargarWhitaker2014}. 
These techniques collectively support concise summaries of atmospheric events by revealing typical geometry, spread, and outliers, and they complement frequency heatmaps introduced in~\cref{sec:method}.

%% file: sec-background.tex
\section{Technical Background: Contour Boxplot}
\label{sec:background}

Given an ensemble of contours, Whitaker et al.~\cite{WhitakerMirzargar2013} introduced contour boxplots to give a statistically grounded, shape-aware summary. A contour boxplot encompasses a representative boundary, central spread, and outliers, analogous to a classic boxplot but tailored to contours.

We first review classic boxplot, and its usage in terms of summarization.
A boxplot is a compact summary for a set of numbers: it shows the median along with range indicators for different percentiles to convey the spread of the set. 
A key step of constructing a boxplot is choosing a sensible center (\ie, the median). 
One way is to define the median by how often a value lies between pairs of other values, which we call the \emph{band depth} of a value in a set.
For 1D data, the band depth of a number is the fraction of all pairwise intervals that contain it; the number with the highest depth is the median. 
For example: for a set $\{2, 4, 5, 7, 12\}$, there are 10 pair intervals. The number 5 lies inside 4 of them, namely $\{2, 7\}$, $\{2, 12\}$, $\{4, 7\}$, and $\{4, 12\}$.
Meanwhile, 4 and 7 lie inside 3 pairs, and 2 and 12 do not lie inside any pair. Thus, the number 5 has the greatest depth and is the median of the set. 

A contour boxplot extends the boxplot idea from numbers to contours. 
Given an ensemble of contours, we assign each contour a band-depth score: for every pair of contours, take the region between them (their \emph{band}); a contour’s depth is the fraction of these pairwise bands that contain it. 
Contours that sit \emph{between} many other pairs are considered to be more central; those that rarely fall between others are considered outliers. 
The contour boxplot encodes the distribution of contour shapes by ordering contours by their \emph{band-depth}: the median contour is the one with the highest depth (most central); a central envelope (e.g., $50\%$) shows where the middle portion of contours typically lies; and an outer envelope (e.g., $100\%$) shows the full spread across all contours, with optional outlier flagged for very low-depth cases. 
In this way, a single figure communicates the typical position of the contour and the variability around it directly in the shape of contours.

\begin{figure}
    \centering
    \includegraphics[width=0.83\columnwidth]{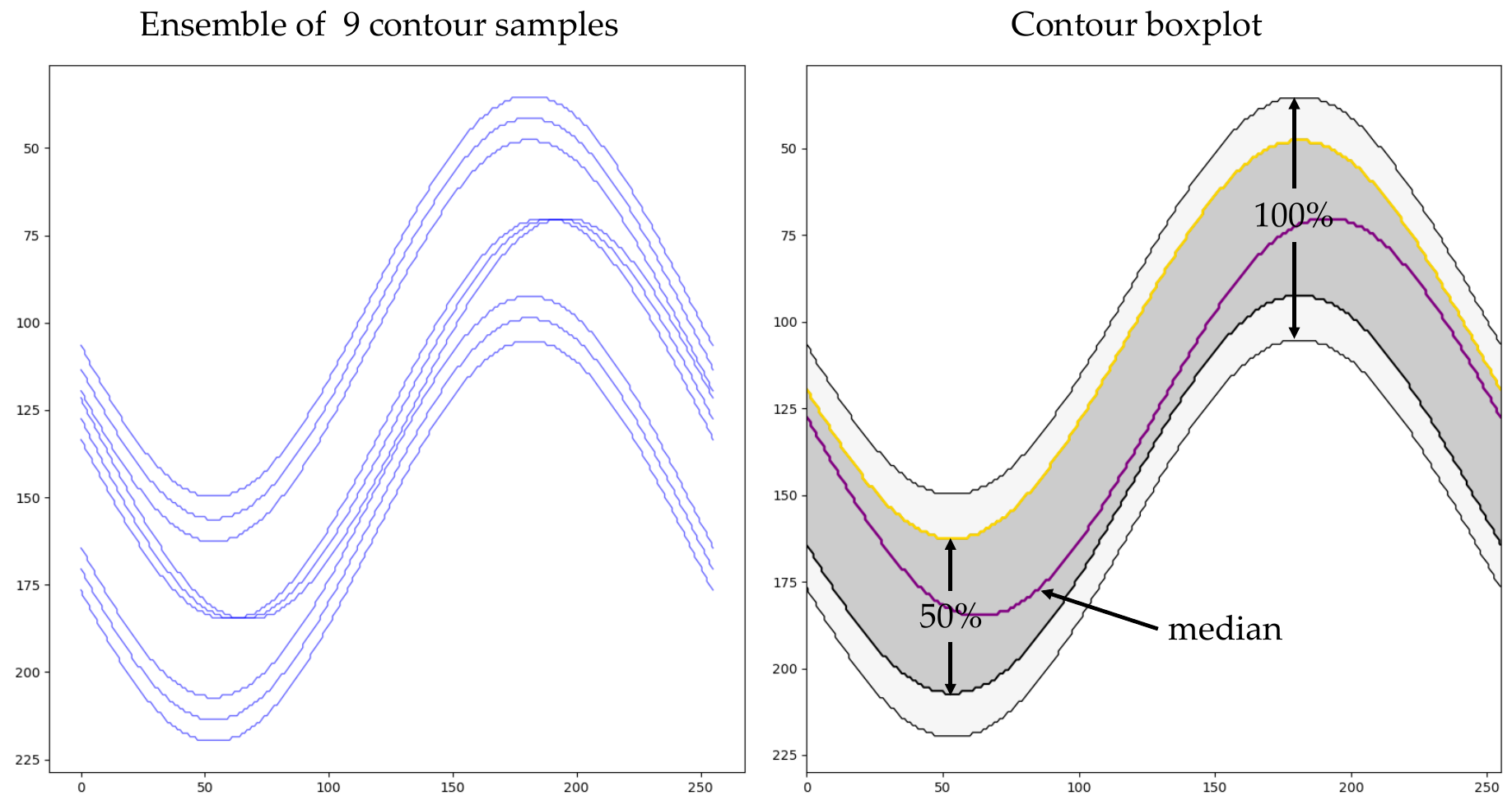}
    \vspace{-2mm}
    \caption{An example of contour boxplot. Left: an ensemble of nine contours generated by sine waves  shifted horizontally and vertically. Right: a contour boxplot showing the median contour (purple), the $50\%$ central envelope (the inner band in gray), and the $100\%$ envelope (the outer band in light grey).}
    \vspace{-3mm}
    \label{fig:background-boxplot}
\end{figure}

We illustrate the concept with a simple ensemble of contours in \cref{fig:background-boxplot}. The left panel shows contours generated from sine waves shifted horizontally and vertically, while the right panel summarizes the ensemble with a contour boxplot. The median curve in purple (i.e., the most central contour by band depth) captures the central tendency. The inner gray band denotes the 50\% central envelope, enclosing the five most central contours, while the outer band marks the 100\% envelope, capturing the full spread of all nine contours and highlighting divergence at peaks and valleys. In short, the median provides a representative shape of the ensemble, while the envelopes convey its variability.

%% file: sec-data.tex
\section{Datasets and Preprocessing}
\label{sec:data}

In this section, we describe the data preparation steps undertaken prior to the detection and characterization of blocking events.

\subsection{Datasets}
\label{sec:description}

We utilize two datasets in our experiments. The first is \textbf{ERA5}, the fifth-generation global reanalysis from the European Centre for Medium-Range Weather Forecasts (ECMWF)~\cite{Hersbach2020}, which serves as a proxy for observational data. ERA5 provides hourly estimates of a wide range of atmospheric state variables, including geopotential, sea-level pressure, and wind speed, at an original spatial resolution of $0.25^\circ \times 0.25^\circ$. For our analysis, we downsample the data to $1^\circ \times 1^\circ$ resolution by averaging over each $4 \times 4$ grid cell block.

The second dataset is derived from a pre-industrial climate simulation using the UK Earth System Model (\textbf{UKESM})~\cite{Sellar2019}, conducted as part of the Coupled Model Intercomparison Project Phase 6 (CMIP6). The UKESM output (version UKESM1-0-LL) has a spatial resolution of $1.25^\circ \times 1.875^\circ$, corresponding to a global grid of $144 \times 192$ cells.

Typical atmospheric blocking events persist for several days; accordingly, we track high-pressure systems over multiple days. To improve robustness against small atmospheric perturbations, we compute \emph{daily averages} for both datasets. For ERA5, we use data from 1979 to 2019, while for UKESM we draw from an arbitrarily defined timeline\footnote{In the simulation, timeline is not real time but a simulation time index.} in the pre-industrial simulations spanning 1960 to 2060, under constant climate forcing conditions.

\para{Ground-truth datasets.}
Thomas et al.~\cite{Nowack2021} created expert-labeled datasets of blocking events in Europe for both ERA5 and UKESM. In this paper, we treat these as ground-truth datasets (GTDs).  
These datasets provide binary labels indicating the presence or absence of blocking events during the June–July–August (JJA) period in a region of Europe ($30-75^\circ$N, $10^\circ$W$-40^\circ$E), following the IPCC AR6 definitions~\cite{Iturbide2020}. The GTDs are constructed following the definition that blocking events persist for at least five consecutive days. The ERA5 GTD spans May 28 to September 4 for each year from 1979 to 2019, whereas the UKESM GTD covers May 27 to September 4 for the years 1960 to 2060, with UKESM employing a standardized 30-day calendar for each month.

\subsection{Data Preprocessing}
\label{sec:preprocessing}
\para{Geopotential height.} 
For both the UKESM and ERA5 datasets, we use the geopotential height at 500 hPa ($Z_{500}$) as a standard variable to characterize atmospheric circulation states. Geopotential height is obtained by dividing the geopotential field from ERA5 by the standard gravitational constant ($\approx 9.80665, \text{m/s}^2$). We focus on the 500 hPa level, a widely used diagnostic for blocking events~\cite{Pinheiro2019,Nowack2021}, as it lies sufficiently above the Earth’s surface to minimize the influence of orographic effects and boundary layer noise, thereby highlighting large-scale circulation patterns.

\para{Geopotential height anomaly.}
To better characterize weather anomalies relative to the expected atmospheric background at any given time of year, we analyze anomalies of geopotential height following previous work~\cite{GrotjahnZhang2017,Pinheiro2019,Nowack2021}. We first compute the \emph{long-term daily mean} (LTDM) of the $Z_{500}$ field at each grid point. Specifically, for each location $(\lat_i, \lon_i)$, we calculate the mean value of $Z_{500}$ for each calendar day across the full dataset period---101 years for UKESM and 41 years for ERA5---referred to as the \emph{daily mean}. This procedure yields a 365-day seasonal cycle at each location. 

\begin{figure}
    \centering
    \vspace{-3mm}
    \includegraphics[width=0.83\columnwidth]{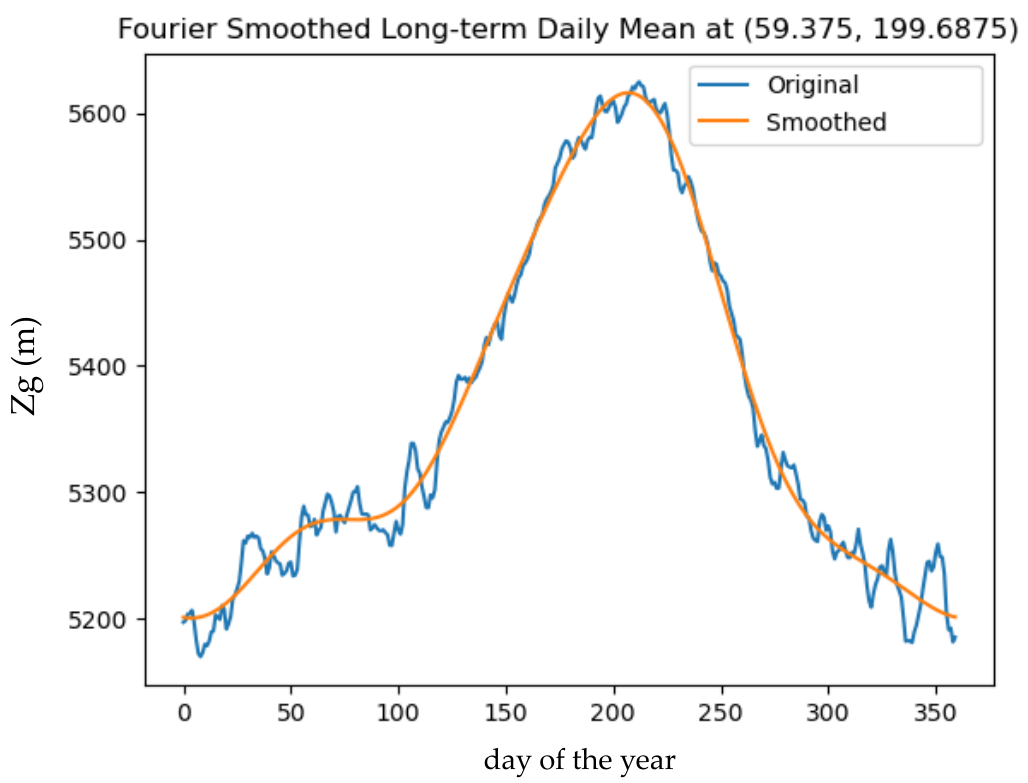}
    \vspace{-2mm}
    \caption{ERA5 dataset: the long-term daily mean (blue) of the geopotential height (Zg, unit: meter or m) at 59.375$\degree$N, 160.3125$\degree$W and its smoothed curve (yellow) by keeping the first six Fourier harmonics.}
    \vspace{-4mm}
    \label{fig:ltdm}
\end{figure}

To smooth the LTDM fields, we approximate this seasonal cycle using the first six Fourier harmonics, thereby reducing noise caused by the limited sample size (e.g., 41 years in ERA5). The result is a smooth climatological background field that we subtract from each daily value to obtain daily $Z_{500}$ anomalies (see~\cref{fig:ltdm} for an example).

Formally, we define the resulting \emph{daily anomaly field} as 
$
Z'_{500} = Z_{500} - \widehat{Z}_{500}^{\text{LTDM}},
$
where $\widehat{Z}_{500}^{\text{LTDM}}$ is the smoothed LTDM field. Additionally, we linearly \emph{detrend} the anomaly fields, following~\cite{Nowack2021}, to remove the influence of background warming in ERA5 and residual model drift in UKESM due to imperfect model spin-up to pre-industrial conditions.  

\para{Normalization of geopotential height anomaly.} 
To make daily anomalies comparable across regions and seasons, we rescale them so that large values are judged relative to the local seasonal variability rather than against a single fixed threshold~\cite{Pinheiro2019}. Starting from the anomaly field $Z'_{500}$ defined above, we compute at each grid cell a \emph{long-term daily standard deviation}, denoted $\widehat{\sigma}_{L}$, from the raw $Z_{500}$ field: for each calendar day, we take the standard deviation across all years and then smooth the resulting 365-day cycle using the first six Fourier harmonics, consistent with the LTDM procedure.

We then normalize as
$
Z^{\text{norm}}_{500} = \frac{Z'_{500}}{\max(100, \widehat{\sigma}_{L})},
$
where the 100 m floor prevents artificially large ratios in regions or seasons with very small background variability (where division by a near-zero standard deviation would otherwise amplify noise and yield spurious exceedances). After this rescaling, values of $Z^{\text{norm}}_{500}$ indicate how unusual a day is for that location and time of year. A candidate high-pressure system is then defined as any contiguous region where $Z^{\text{norm}}_{500} \geq \lambda$, allowing a single threshold $\lambda$ to be applied consistently across regions and seasons.

%% file: sec-method.tex
\section{Method}
\label{sec:method}
This section outlines our pipeline. We begin by tracking connected high-pressure systems using a geometry-based criterion (\cref{sec:tracking}). Next, we identify blocking events as systems that persist for at least five consecutive days (\cref{sec:detecting}), with parameter tuning described in \cref{sec:implementation}. Finally, we introduce our main contribution: an uncertainty visualization framework for characterizing the spatiotemporal patterns of blocking events (\cref{sec:uncertainty}).

\subsection{Tracking High-Pressure Systems}
\label{sec:tracking}

\para{Extracting high-pressure systems.}
Mid-latitude blocking is typically associated with quasi-stationary high-pressure systems in the normalized 500\,hPa geopotential-height anomaly field \(Z^{\text{norm}}_{500}\).
For each day \(t\) (restricted to the European domain), we form the superlevel set $S_\lambda(t)\;=\;\{\,x \mid Z^{\text{norm}}_{500}(x,t)\ge\lambda\,\}.$
A \emph{superlevel-set component} is defined as a contiguous region within this set, representing the footprint of a candidate high-pressure system on that day.

\para{Geometry-based tracking.}
Let $A\subseteq S_\lambda(t)$ and $B\subseteq S_\lambda(t{+}1)$ be components on consecutive days. We measure their (latitude-weighted) area overlap by
$\Omega(A,B)\;=\;\sum_{x\in A\cap B} w(x),$ where 
$w(x)=\cos(\text{lat}(x))$.
We connect $A$  to $B$  if $\Omega(A,B)\ge C$ for some threshold $C$. Processing days sequentially yields trajectories that capture quasi-stationary evolution while permitting modest day-to-day displacements. Default values of parameters are reported in \cref{sec:implementation}.

\para{Ambiguity in system correspondences.} 
Because systems may appear, disappear, merge, or split, correspondences need not be unique. We retain all trajectories satisfying $\Omega\!\ge C$; these plausible continuations are carried forward to the persistence test below.

\subsection{Detecting Blocking Events}
\label{sec:detecting}
\para{Persistence test and labels.}
Following Thomas et~al.~\cite{Nowack2021}, a \emph{blocking event} is present when a tracked system persists for at least five consecutive days. Let \(\mathcal{T}\) be the set of trajectories constructed above, and let $D(\tau)$ denote the set of calendar days covered by a trajectory \(\tau\in\mathcal{T}\). We label a day \(t\) as \emph{positive} if
\[
\exists\,\tau\in\mathcal{T}\ \text{such that}\ \bigl|D(\tau)\bigr|\ge 5
\quad\text{and}\quad
t\in D(\tau).
\]
That is, \emph{all} days covered by any trajectory whose duration is at least five consecutive days receive a positive label. Equivalently, whenever a trajectory contains any 5-day contiguous span, every day on that trajectory is marked positive; e.g., a system lasting July~1--8 renders July~1--8 positive.

\para{Robustness to merges and splits.}
Because high-pressure systems may appear, disappear, merge, or split, multiple trajectories can interact. We define a blocking event as any trajectory that remains continuous over a five-day window, without enforcing unique correspondences between systems. This approach preserves robustness to minor boundary changes, while the overlap threshold $C$ enforces day-to-day quasi-stationarity.
All the positively labeled components constitute our ensemble of \emph{blocking footprints}; their boundaries are the \emph{blocking boundaries} used in \cref{sec:uncertainty}.

\subsection{Uncertainty Visualization}
\label{sec:uncertainty}
While the tracking of high-pressure systems is relatively standard, relying on a region-overlap criterion, the detection of blocking events requires careful parameter tuning (as detailed in \cref{sec:implementation}). Our primary contribution is an uncertainty visualization framework for characterizing the spatiotemporal patterns of blocking events. 

Blocking footprints can vary across days, months, and years, even on the same calendar date. To convey their typical locations and variations over time, we summarize ensembles of blocking events using three uncertainty-aware visual encodings:
\begin{itemize}[noitemsep,leftmargin=*]
\item \textbf{Contour boxplots} that capture blocking boundary variations;
\item \textbf{Frequency heatmaps} that encode occurrences;
\item \textbf{3D temporal stacks} that situate these patterns in time. 
\end{itemize}

To capture the frequency, duration, and intensity of blocking events across different spatiotemporal scales, we construct ensembles of varying types for each dataset. In the ERA5 dataset spanning 41 years, for example, a fixed calendar day (e.g., July 4) defines a \emph{daily ensemble} that aggregates all blocking events detected on that day across the 41-year period. Likewise, a fixed calendar month (e.g., July) defines a \emph{monthly ensemble} that collects all blocking events detected throughout that month across the 41 years. Similarly, a summer season (e.g., June, July, and August) gives rise to a \emph{seasonal ensemble} that contains all blocking events detected throughout the season in the 41-year period. 

\para{Contour boxplots.}
Given an ensemble of blocking events, we collect all blocking boundaries into a single set of contours $\mathcal{G}=\{\gamma_1,\dots,\gamma_n\}$. We then compute:
\begin{enumerate}[noitemsep,leftmargin=*]
\item \textbf{Pairwise bands.} For each contour $\gamma_i$, iterate over all unordered pairs $(\gamma_j,\gamma_k)$ for all $j<k$, and form the \emph{band} between $\gamma_j$ and $\gamma_k$, that is, the region lying between the two blocking boundaries within the domain.
\item \textbf{Mismatch matrix.} For each band, measure the fraction of $\gamma_i$ that lies outside that band, and store this information in a mismatch matrix $M$ (with elements valued within $[0,1]$).
\item \textbf{Relaxed depth with $\varepsilon$.} 
We treat $\gamma_i$ as \emph{inside} a band if its mismatch for that pair is $\leq \varepsilon$. The \emph{relaxed band-depth score} $D_\varepsilon(\gamma_i)$ is defined as the fraction of all pairs for which $\gamma_i$ is labeled as \emph{inside}. We then sweep over a small grid of $\varepsilon$ values and select the smallest $\varepsilon$ that stabilizes the ranking of the most central contours and produces a well-defined central region, as determined by the values of $D_\varepsilon$.
\item \textbf{Contour boxplot.} The following three elements constitute the contour boxplot we display:
\begin{itemize}[noitemsep,leftmargin=0pt]
\item  The \textbf{median} is the contour with the largest relaxed band-depth $D_{\varepsilon}$. 
\item  The \textbf{50\% envelope} is the union-minus-intersection of the deepest half of contours quantified by $D_{\varepsilon}$. 
\item  The \textbf{100\% envelope} is the union-minus-intersection of all contours. 
\end{itemize}
\end{enumerate}
See~\cite{WhitakerMirzargar2013} for more details on implementing the contour boxplot, including tuning the parameter $\varepsilon$.

\para{Frequency heatmaps.}
For a given ensemble of blocking events, we construct a frequency heatmap by counting, at each grid cell, the number of days in which that cell lies within a detected blocking footprint. The resulting heatmap encodes the frequency of blocking events across the ensemble, highlighting where blocking events are most likely to occur within a daily or monthly ensemble.

The frequency map highlights where blocking events repeatedly occur and how concentrated they are, while the contour boxplot summarizes the \emph{representative} boundary (median) and the \emph{spread} of boundaries (50\% and 100\% envelopes). Taken together, the median provides a representative footprint, the 50\% envelope identifies the common core, the 100\% envelope captures the full spatial extent observed in the ensemble, and the frequency heatmap indicates the likelihood of blocking events.

\begin{figure}[!ht]
\centering
\includegraphics[width=1.0\columnwidth]{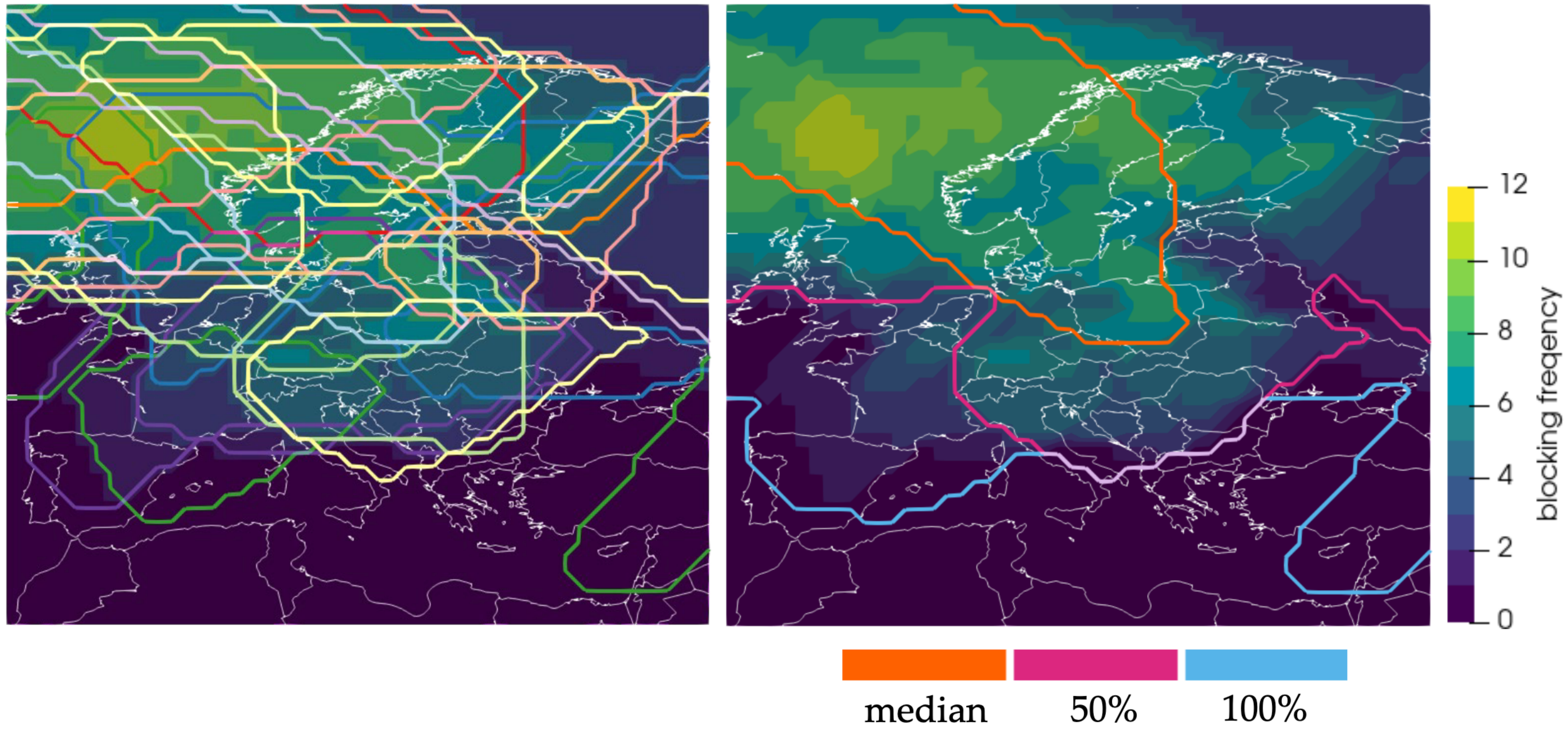}
\vspace{-6mm}
\caption{ERA5 dataset (1979–2019), June 5 daily ensemble. Left: the boundaries of 23 ensemble members, which are blocking events detected across 41 years. Right: the contour boxplot of the ensemble, including the median blocking boundary (orange), 50-percentile central envelope (meganta), and 100-percentile envelope (blue). The background frequency heatmap highlights blocking frequency.}
\vspace{-5mm}
\label{fig:boxplot-sample}
\end{figure}

As a concrete example, \cref{fig:boxplot-sample} summarizes the daily ensemble for June 5 in the ERA5 dataset (1979–2019). In 23 of the 41 years, June 5 was classified as a (positive) blocking day. The frequency heatmap shows, across those 23 years, how often each location fell within a detected blocking footprint. The left panel overlays the 23 extracted block boundaries, while the right panel summarizes them with a contour boxplot. In this boxplot, the median contour is the most central member of the ensemble and serves as the representative. The 50\% contour encloses the central half of all boundaries, while the 100\% contour traces the union of all boundaries. For the June-5 daily ensemble, the 100\% contour delineates the full spatial extent of detected blocking events, with its southern edge marking the farthest south any June-5 blocking event has reached.

\para{3D temporal stacks.}
As shown in \cref{fig:boxplot-sample}, a contour boxplot combined with a frequency heatmap summarizes the variability of blocking events for a single calendar day across 41 years, but it does not capture how their shapes and occurrences evolve throughout the season. To explicitly incorporate temporal information, we construct two complementary 3D temporal stacks along the time dimension for a given ensemble: the \emph{3D median stack} and the \emph{3D frequency stack}.

To construct the 3D median stack, we extract the median contour from each day’s contour boxplot (the representative boundary) and place it in a 3D scene, with latitude and longitude on the horizontal axes and calendar time on the vertical axis. Stacking these daily medians produces a filament-like sketch of how a representative footprint shifts and persists over time. To reduce visual clutter, we restrict the stack to median contours without envelopes.

To construct the 3D frequency stack, we compute a frequency heatmap for each day and stack them along the same time axis to form a 3D volume, where higher values indicate dates and locations with more frequent blocking events. 

\begin{figure}[!ht]
    \centering
    \vspace{-2mm}
    \includegraphics[width=0.9\columnwidth]{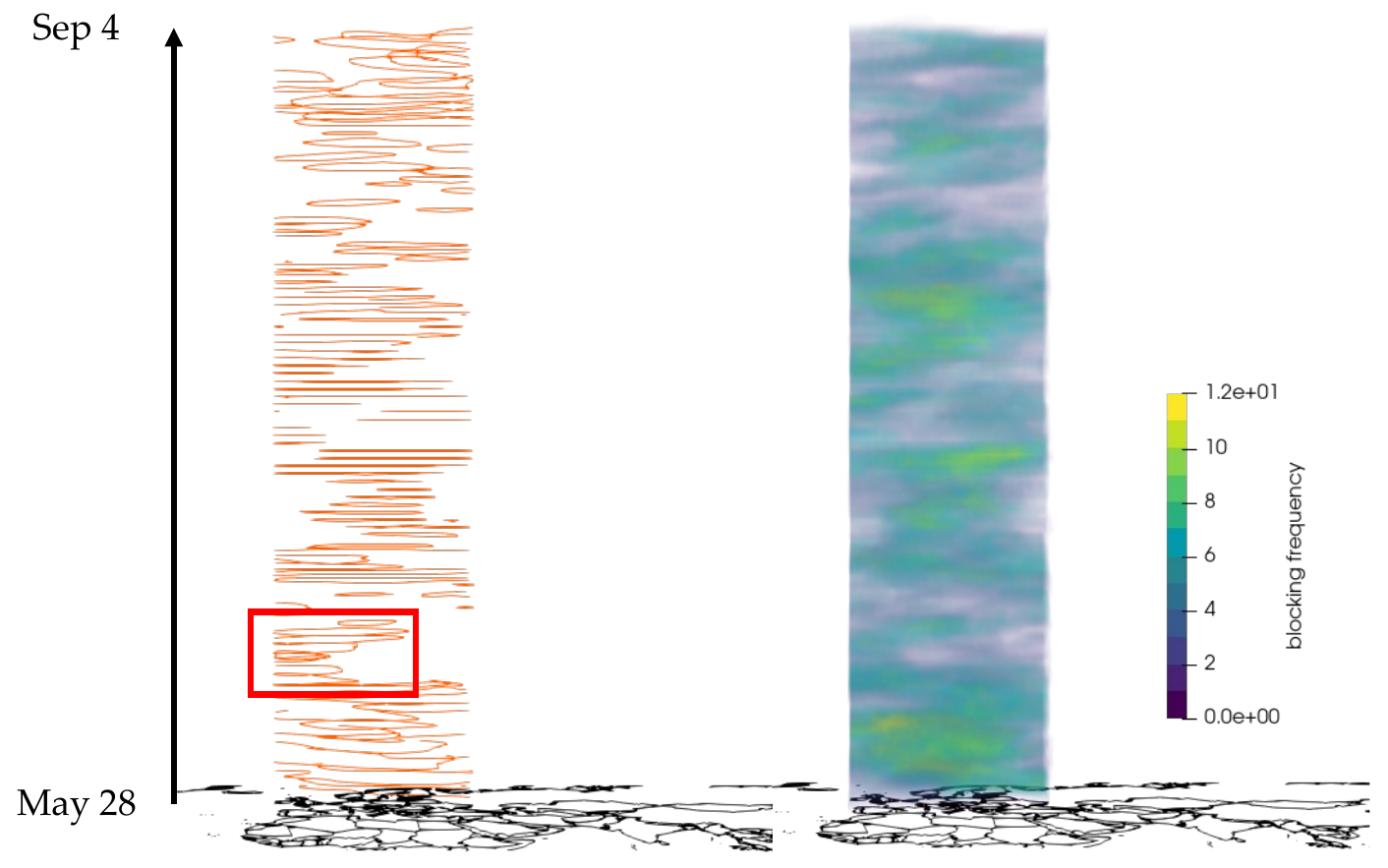}
    \vspace{-2mm}
    \caption{3D temporal stacks of a seasonal ensemble (ERA5 dataset, May 28–Sep 4, 1979–2019). Left: a 3D median stack, where daily median contours are stacked by date along the $z$-axis (earlier dates at the bottom). Right: a 3D frequency stack, constructed by stacking daily frequency heatmaps into a 3D volume along the same $z$-axis, with color encoding frequency. 
    The red box highlights a mid-June interval during which the median contours cluster over the west of Europe.
    Together, these 3D temporal stacks summarize the evolution and spatial persistence of blocking events throughout the season.}
    \vspace{-2mm}
    \label{fig:boxplot-stack}
\end{figure}

The two 3D temporal stacks provide a spatiotemporal overview that complements the 2D contour plots and frequency heatmaps. In the 3D median stack, clusters of vertically aligned median contours indicate multi-day concentrations of blocking events. In the 3D frequency stack, bright regions highlight periods and locations with frequent blocking events. These 3D temporal stacks are illustrated in \cref{fig:boxplot-stack}. 

%% file: sec-implementation.tex
\section{Implementation}
\label{sec:implementation}

Our implementation is available at~\url{https://github.com/tdavislab/atmos-blocking-uncertainty}. 

\para{Configurations.}~All experiments are conducted on a laptop with a 12th Gen Intel(R) Core(TM) i9-12900H 2.50 GHz CPU with 32 GB memory. 
Data processing, tracking, evaluation, and visualizations are all implemented in Python.
We use the \emph{ParaView 5.13.3}~\cite{AhrensGeveciLaw2005} for the frequency heatmap and volume rendering of the 3D frequency stack. 

\para{Pipeline overview.}
According to \cref{sec:method}, we first track high-pressure systems directly in the daily normalized 500 hPa geopotential-height anomaly field, $Z^{\text{norm}}_{500}(x,t)$. 
For each JJA day $t$, we threshold at level $\lambda$ and take the connected components of the superlevel set $\{Z^{\text{norm}}_{500} \geq \lambda \}$ as system candidates. 
To track systems across time, we employ a region-overlap strategy by counting intersecting pixels weighted by their latitudes $w=\cos(\text{lat}(x))$, where $\text{lat}(x)$ is the latitude of the pixel $x$.
Pairs of systems whose weighted overlap exceeds a threshold $C$ are connected and thus producing time-ordered trajectories as days are processed sequentially. 
A day is labeled \emph{positive} if it belongs to any trajectory lasting at least five consecutive days.

\para{Parameter tuning.}
We tune two hyperparameters: the superlevel-set threshold $\lambda$ and the overlap threshold $C$. Following DG83~\cite{DG83}, which sets $\lambda=1.5$ for normalized $Z_{500}$, we conduct an exhaustive grid search with $\lambda \in \{1.0,1.1,\dots,2.0\}$ and $C \in \{5,6,\dots,40\}$. For each dataset, the first half of the years is used as a tuning set for five-fold cross-validation: the years are partitioned into five folds, each parameter pair $(\lambda,C)$ is scored by training on four folds and validating on the held-out fold, and the pair maximizing the mean F1-score across folds is selected. The chosen $(\lambda,C)$ is then fixed for all subsequent analyses.
Because our detector is a transparent threshold-and-tracking procedure (i.e., a white-box index), we follow common practice in the blocking-index literature and report results on the full dataset rather than restricting to a separate test set~\cite{Nowack2021}. 
The optimal set of parameters for the two datasets is
\begin{itemize}[noitemsep]
    \item ERA5: $\lambda=1.2$, $C=31$
    \item UKESM: $\lambda=1.0$, $C=31$
\end{itemize}

%% file: sec-results.tex
\section{Experimental Results}
\label{sec:results}

To demonstrate the utility of our framework, we first evaluate its performance in detecting blocking events, compared against two baseline methods: SOM-BI~\cite{Nowack2021} and DG83~\cite{DG83,Pinheiro2019}. The baselines are described in \cref{sec:baseline}, and the evaluation results are presented in \cref{sec:detection}. We then provide a case study of the 2003 European heatwave in \cref{sec:case-study}, showcasing our uncertainty visualizations. Finally, in \cref{sec:result-uncertainty}, we summarize the spatiotemporal patterns of blocking events and visually analyze their variations. 

\subsection{Baseline Detection Methods}
\label{sec:baseline}

Both DG83~\cite{DG83,Pinheiro2019} and SOM-BI~\cite{Nowack2021} detect blocking events from persistent anomalies in the 500 hPa geopotential height field ($Z_{500}$).

\para{DG83.} 
In the original work of Dole and Gordon~\cite{DG83}, $Z_{500}$ anomalies are computed as departures from a long-term seasonal mean and rescaled by a latitude factor $(\sin 45^\circ / \sin \phi)$ ($\phi$: latitude) to account for the meridional variation of planetary vorticity. A grid point is considered \emph{blocked} (i.e., part of a blocking footprint) if the scaled anomaly exceeds a fixed threshold (originally $\geq 100$ m) for a sustained duration (originally $\geq 10$ days).

For global intercomparison, Pinheiro et al.~\cite{Pinheiro2019} modified DG83 by replacing the fixed 100 m threshold with a spatiotemporally varying threshold of $1.5\sigma$ (with a floor of 100 m), where $\sigma$ is the smoothed long-term standard deviation of geopotential height at each grid point. Blocking events are matched across days when they overlap in at least one pixel and satisfy a duration constraint of $\geq 5$ days; we adopt this modified DG83 by Pinheiro et al. as our baseline for comparison.

\para{SOM-BI.}
SOM-BI~\cite{Nowack2021} is a state-of-the-art blocking index based on geopotential height. It trains a self-organizing map (SOM) on daily $Z_{500}$ anomaly maps over Europe to learn a compact set of representative circulation patterns, referred to as SOM nodes. During training, each day is assigned to its best-matched node, and the matched node together with its neighbors on the SOM grid are updated toward that day’s field using a decaying learning rate and neighborhood size. This process is iterated until the node patterns converge, after which each day is mapped to its best-matched node.

SOM-BI then applies a 5-day sliding window over the dataset, treating each sequence of best-matched nodes as a candidate \emph{blocking signature}. Using ground-truth labels of blocked 5-day periods, all distinct signatures are scored, and the subset that maximizes the F1-score is retained. A day is labeled to be \emph{blocked} (i.e., exhibiting blocking event) if any 5-day window containing it matches one of the retained signatures. 

\para{Comparison with our approach.}
Both DG83 and our detection method apply a superlevel-set threshold to the normalized $Z_{500}$ anomaly field, whereas SOM-BI employs an unsupervised pattern-classification framework based on a self-organizing map~\cite{DG83,Nowack2021}. The key difference is that our method explicitly enforces spatial stationarity: components from consecutive days must overlap by an area greater than $C$, ensuring that only slowly evolving, spatially coherent high-pressure systems are linked into a trajectory and counted toward the multi-day lifetime requirement. By contrast, DG83 requires only minimal day-to-day overlap (as little as a single pixel) between footprints. As a result, rapidly translating highs that do not constitute a blocking event may still satisfy DG83’s duration rule, whereas our area threshold $C$ enforces quasi-stationarity.

To maintain precision, DG83 typically relies on a relatively high anomaly threshold (e.g., $1.5\sigma$), which helps reduce false positives but also makes the method less sensitive to moderate events. In contrast, our overlap criterion tolerates modest spatial displacements while still enforcing quasi-stationarity, thereby allowing the use of a lower $\lambda$ without inflating noise.

On the other hand, SOM-BI differs in spirit: it first learns the representative of daily $Z_{500}$ anomaly patterns and then labels days via the best-performing 5-day pattern sequences against the ground truth~\cite{Nowack2021}. 
This can be robust to synoptic variability, but the classification hinges on SOM design and training labels. 
Besides, SOM-BI yields only binary labels rather than explicit boundaries of high-pressure systems. 
In contrast, our tracking directly produces contiguous footprints each day, which we use consistently for case studies involving contour boxplots and frequency heatmaps. 

\subsection{Blocking Event Detection Evaluation}
\label{sec:detection}
\input{tab-detection-gtd}
\para{Evaluation with ground-truth labels.}
We assess all blocking event detection methods against the ground-truth (GTD) labels. 
For DG83 and our detection method, scores are computed from our own implementation. 
For SOM-BI, we report the published results of Thomas et al.~\cite{Nowack2021}, obtained on the same datasets and timespan.

We evaluate day by day against the ground-truth labels using \emph{accuracy}, \emph{precision}, \emph{recall}, and \emph{F1}-score. 
A ``positive'' label means \emph{blocked} and a ``negative'' label means \emph{not blocked}. 
A day is a true positive (TP) if both the output label and GTD are positive; a true negative (TN) if both are negative; a false positive (FP) if the tool labels it positive but GTD is negative; and a false negative (FN) if the tool labels it negative but GTD is positive.

\para{Detection summary.}
Standard confusion metrics are computed over the evaluation period, including accuracy, precision, recall and F1-score. These metrics are reported in \cref{tab:detection} for ERA5 (reanalysis) and UKESM (simulation) datasets across the three detectors (DG83, SOM-BI, and ours). 
For \textbf{ERA5}, all methods achieve comparable accuracy and precision, but our method attains the highest recall and therefore the top F1-score. 
For \textbf{UKESM}, our method performs consistently strongest overall: its precision is on par with DG83, while recall is substantially higher, yielding a higher F1-score. 

This result reflects our design choices: unlike DG83, which relies on a relatively high anomaly threshold to safeguard precision, our method enforces quasi-stationarity through a region-overlap constraint, filtering out fast-moving high-pressure systems before they can contribute to false positives. This constraint enables the use of a more permissive anomaly threshold without sacrificing precision (ERA5: $\lambda=1.2$ vs. 1.5 in DG83; UKESM: $\lambda=1.0$ vs. 1.5), while recovering additional modest events and improving recall. 

Relative to SOM-BI, we achieve similar or higher F1-score on both datasets and, moreover, produce day-by-day contiguous footprints of the high-pressure systems, which feed directly into our downstream analysis.
In comparison, SOM-BI yields a binary time series and typically relies on composite of additional heatmaps (such as $Z_{500}$, vertically integrated potential vorticity) for spatial context.

\para{Disagreement with DG83.}
Because both DG83 and our method apply thresholds to the normalized $Z_{500}$ field, we examine cases where their GTD-based labels disagree (\cref{tab:improvement}). With a more lenient normalized anomaly threshold $\lambda$, our method correctly identifies 161 additional blocked days in ERA5 and 440 in UKESM that DG83 misses, at the cost of introducing more false positives (161 in ERA5 and 340 in UKESM) that DG83 correctly marks as unblocked. This reflects the standard precision-recall trade-off: lowering $\lambda$ increases sensitivity (recall) but can also admit marginal cases.

Two design choices temper this loss of precision in our framework: the region-overlap requirement $C$ and trajectory building, which filter out fast-moving highs and favor quasi-stationary systems. Consistent with this, DG83 flags only 14 (ERA5) and 109 (UKESM) blocked days that our method does not, which are likely intensified systems that failed our stationarity and lifetime criteria. In contrast, our method correctly screens out 100 (ERA5) and 305 (UKESM) days as unblocked that DG83 labels as blocked.

Overall, relative to DG83, our approach shifts the balance toward higher recall for modest or slightly drifting events, while maintaining competitive precision through the overlap threshold $C$.

\para{Prevalence and bias.}
While our precision remains comparable to the baselines, the number of days labeled as blocked by our detector is slightly higher than the GTD base rate in ERA5 (37.3\% vs. 33.4\%), indicating a mild positive bias toward detection. 
In UKESM, our prevalence closely matches GTD (29.4\% vs. 29.0\%), whereas DG83 under-detects (25.9\%). 
Taken together, these results suggest our method trades a small increase in prevalence for meaningful gains in recall, with precision preserved by the explicit overlap-based stationarity check.

\para{Potential improvement.}
Among the three blocking event detection tools, only our method reaches a considerable difference between the precision and recall on the ERA5 dataset, indicating bias towards over-detection than under-detection. This is because our parameter tuning process prioritizes maximizing the F1-score on the validation data without much consideration about the difference between precision and recall. 
We may improve the parameter tuning process to find the parameter setting that minimizes the difference between precision and recall to better balance the ratio of detected blocked days.

\input{tab-dg83-alignment}

\para{Out-of-season agreement.}
Although $\lambda$ and $C$ are tuned using only JJA (summer) cases, we also test generalization to the rest of the year. 
To mitigate seasonal amplitude differences, we pre-process $Z_{500}$ anomalies using the global/seasonal normalization introduced by Pinheiro et al.~\cite{Pinheiro2019}, and we compare our monthly outputs against DG83 implemented with the same normalization; see~\cref{sec:preprocessing} for the normalization details. 
\cref{tab:agreement} reports monthly precision, recall, and F1-score for this pairwise agreement.

For \textbf{ERA5}, within JJA (the tuning season), the F1-score agreement with DG83 is 0.885 (June), 0.855 (July), and 0.806 (August). 
Outside JJA, the lowest monthly F1-score is 0.777 (November), with the next-lowest 0.840 (January).
All values are comparable to the in-season range, indicating broadly similar alignment across months, with November showing the largest divergence.

For \textbf{UKESM}, a similar pattern holds: in JJA the F1-scores are 0.756 (June), 0.812 (July), and 0.802 (August); 
outside JJA, the lowest F1-score occurs in October (0.710), followed by May (0.780). 
Across all the months aside from October, there are no pronounced spikes or dips in F1-score.
Overall, the absence of strong seasonal variations in monthly F1-score suggests that, with the same normalization applied, our method maintains a generally consistent level of agreement with DG83 beyond the season used for tuning.

\begin{figure*}[!ht]
    \centering
    \includegraphics[width=1.68\columnwidth]{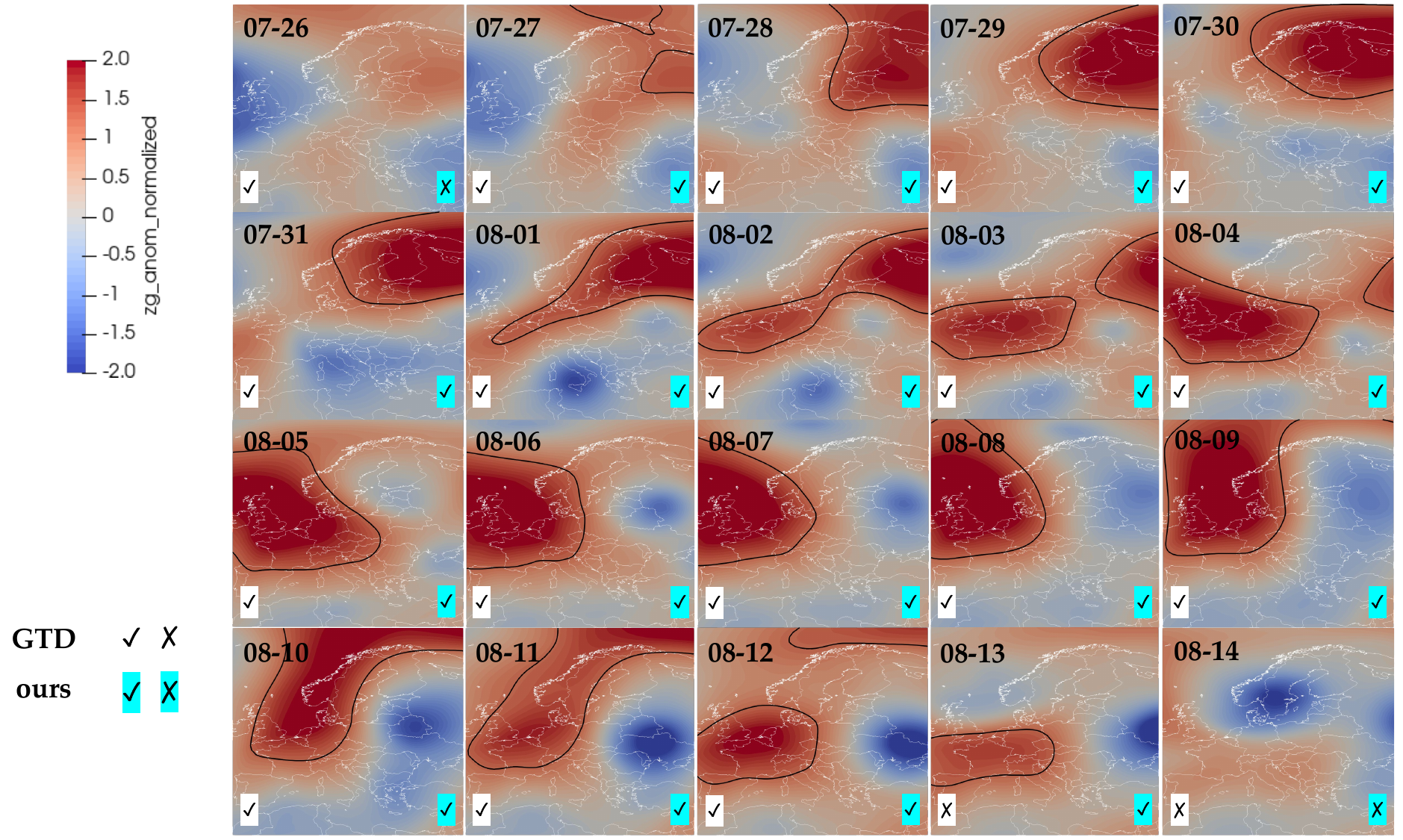}
    \vspace{-2mm}
\caption{ERA5 dataset: the frequency heatmap of normalized $Z_{500}$ anomalies over Europe for July 26 - August 14, 2003. White lines show national borders and coastlines. The black contour (isovalue=1.2) delineates the high-pressure boundary as defined by our detection method. Bottom annotations in each map report our detection and the ground-truth labels (GTD) for each date; tick = blocked, cross = not blocked.} 
    \label{fig:case-study-2003}
    \vspace{-4mm}
\end{figure*}

\subsection{Case Study: 2003 European Heatwave}
\label{sec:case-study}

The 2003 European heatwave is a canonical benchmark for blocking event detection: from late July into mid-August, a persistent, quasi-stationary anticyclonic circulation settled over Western Europe, displacing the North Atlantic jet poleward, suppressing precipitation, and favoring clear-sky subsidence and extreme near-surface warming~\cite{BlackBlackburn2004}. 
According to the record, the most intense phase occurred around August 6-12 as the anticyclonic block centered near France, with $Z_{500}$ anomalies and stagnant flow maintaining accumulated heat over the region~\cite{LiuHe2020,TrigoGarcia2005,GarciaDiaz2005}. 
The summer-mean temperature anomalies exceeded $3\degree$C over broad areas and locally reached several standard deviations above the 1961–1990 baseline~\cite{FischerSeneviratne2007,Schar2004}. 
The societal toll was enormous: more than 70,000 people died across Europe, with France particularly affected~\cite{RobineCheung2008}.
As a case study, this heatwave provides a well-documented sequence of onset, maintenance, and decay of a high-impact blocking event against which to evaluate both detection skill and attribution of sustained heat extremes.

\para{Case-study detection.}
\cref{fig:case-study-2003} shows normalized $Z_{500}$ anomalies from the ERA5 observations, with a black contour (isovalue $1.2$) marking the diagnosed high-pressure domain from July 26 to August 14, 2003.
From July 27 through August 13, every day is detected as ``blocked'' by our method. 
The detected high-pressure system evolves from an initial peak over northeastern Europe on July 27–28 to a westward-expanded ridge that covers France, Germany, the U.K., and the adjacent North Atlantic by August 4–12. 
The system then weakens on August 13 and dissipates on August 14. 

This day-by-day evolution aligns with independent analyses for the 2003 European heatwave, which document exceptionally large positive $Z_{500}$ anomalies over western Europe during early-mid August and a subsequent decline thereafter~\cite{BlackBlackburn2004,TrigoGarcia2005}.
Importantly, the signal we highlight with contours (anomalously high $Z_{500}$) captures the same large, quasi-stationary high-pressure pattern that defines a blocking event in practice. 
When this positive anomaly spreads over a wide area and lasts for several days, winds weaken and the air mass stagnates, allowing heat to build. 
In the 2003 heatwave case, our uninterrupted run of blocked days aligns with the observed intensification and lifetime of extreme heat over France and neighboring regions in early-mid August~\cite{BlackBlackburn2004,TrigoGarcia2005}.

\para{Evaluation.}
We compare our blocking event detections with the ground truth labels. Only two dates disagree: July 26 and August 13. 
On July 26, our method finds no grid cells above the scalar threshold, whereas the GTD marks the day as blocked; on August 13, a pronounced high-pressure anomaly remains over western Europe, yet the GTD is marked as unblocked. 
Given the judgment involved in expert ground-truth labeling, minor discrepancies are unsurprising. For example, on July 26 (\cref{fig:case-study-2003}, top-left), the normalized anomaly falls just below our threshold, yet a nascent positive signal is evident and subsequently intensifies into the high-pressure system observed the following day.

\begin{figure}[!ht]
    \centering
    \vspace{-2mm}
    \includegraphics[width=0.9\columnwidth]{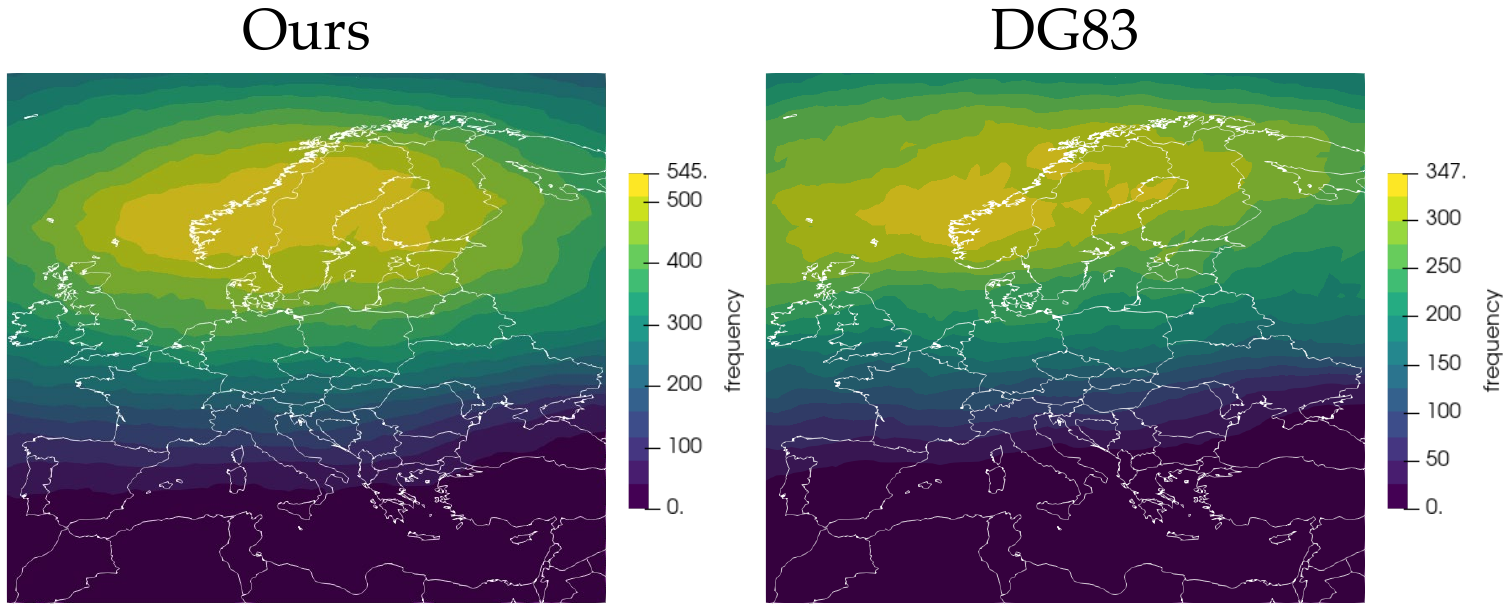}
    \vspace{-2mm}
    \caption{ERA5 dataset (May 28-Sep 4, 1979-2019). The spatial distribution of blocking event frequency over Europe.}
    \vspace{-4mm}
    \label{fig:era5-frequency}
\end{figure}

\begin{figure*}[!ht]
    \centering
    \includegraphics[width=1.6\columnwidth]{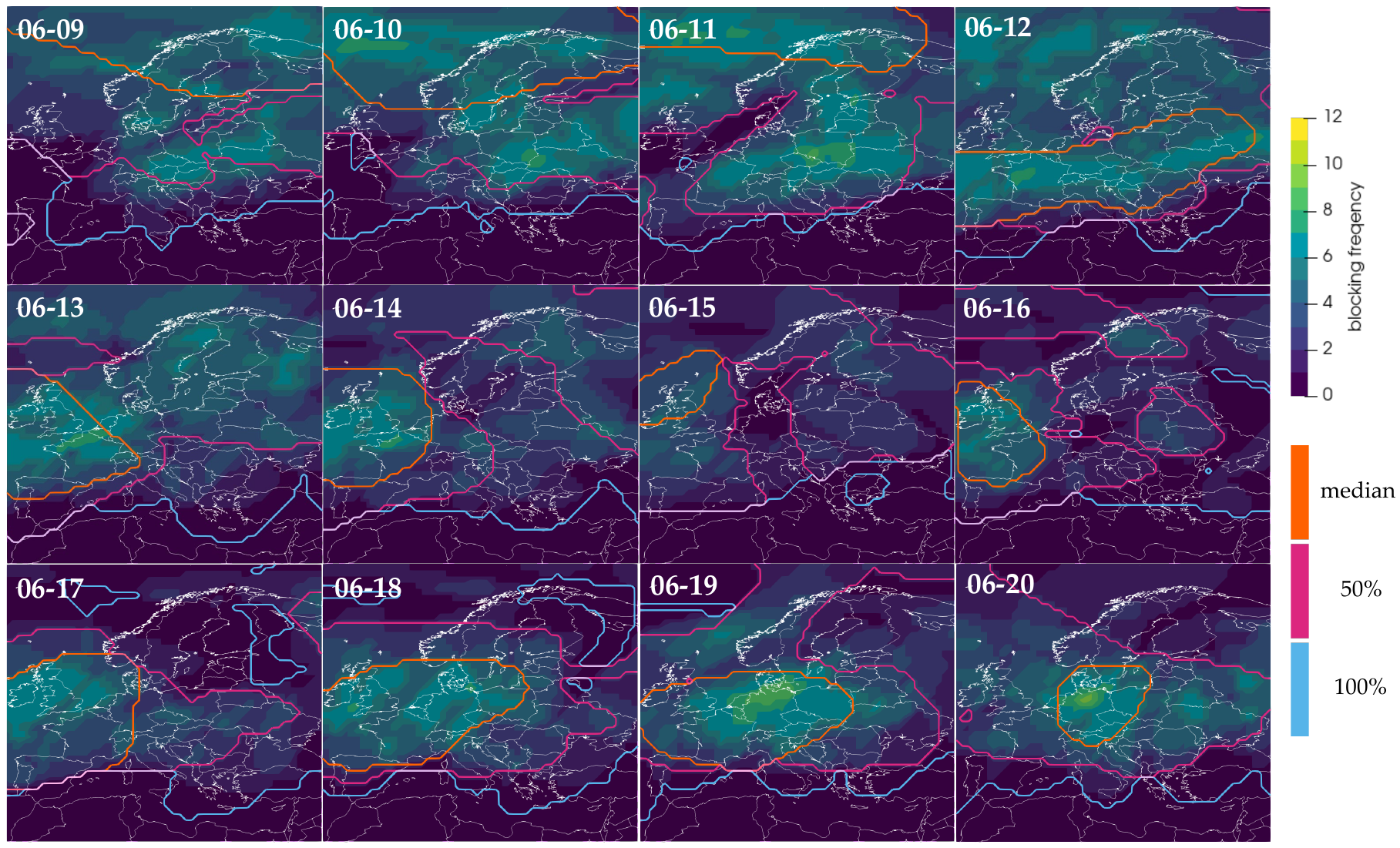}
    \vspace{-3mm}
    \caption{Daily contour boxplots for June 9-20 (ERA5, 1979-2019). In each panel, the background frequency heatmap shows blocking frequency across the analysis period; overlaid contour boxplots summarize the ensemble of detected blocking boundaries for that date: the median boundary, the 50-percentile  envelope, and the 100-percentile envelope. The contributing years vary by day, since not every year has a blocking event on each date.} 
    \label{fig:boxplot-daily}
    \vspace{-3mm}
\end{figure*}

\begin{figure*}[!ht]
    \centering
    \includegraphics[width=1.6\columnwidth]{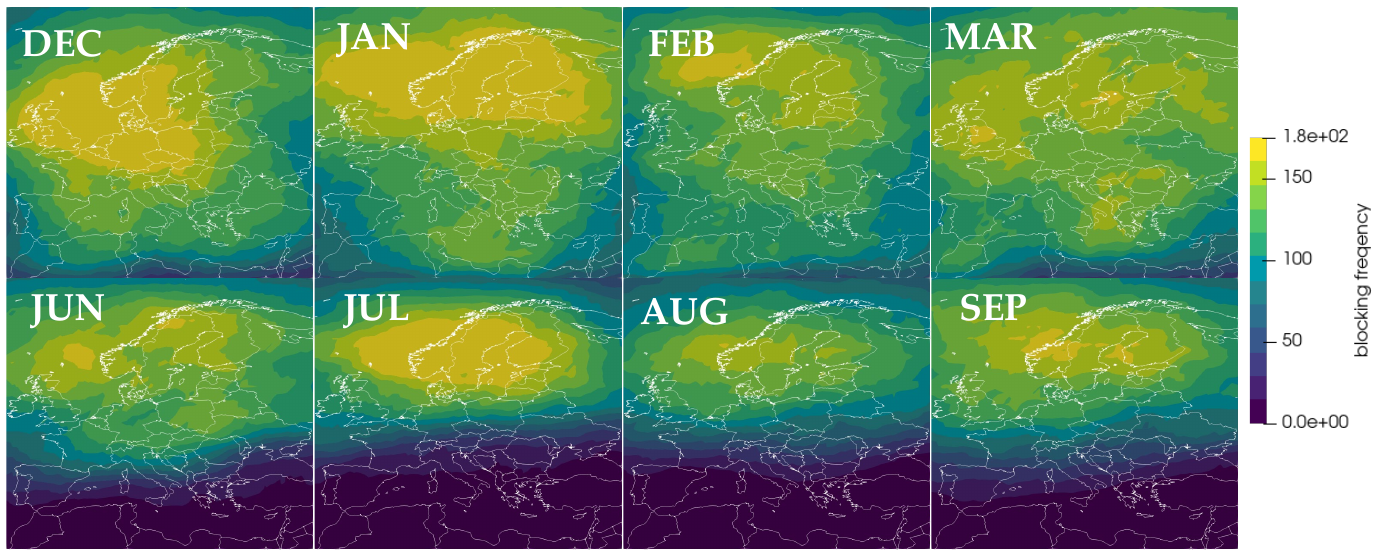}
    \vspace{-3mm}
    \caption{Frequency heatmaps for a monthly ensemble (ERA5, 1979–2019). A $2\times 4$ grid of heatmaps: top row includes the winter months (December, January, February, and March); bottom row includes the summer months (June, July, August, and September). For each month, the heatmap shows the frequency of blocked days aggregated across the 41-year dataset (using the same threshold and domain as in the case studies). 
    Contour boxplots are omitted here: monthly ensembles contain many more events than a single day, making their computation expensive.
    }
    \vspace{-5mm}
    \label{fig:boxplot-monthly}
\end{figure*}

\subsection{Uncertainty Analysis and Visualization}
\label{sec:result-uncertainty}

Our detection method not only identifies days with blocking events but also depicts the boundary of the high-pressure systems associated with them.
Based on our detection results, we present an extensive analysis of the spatiotemporal distribution of blocking events in Europe.
We focus our analysis on the ERA5 dataset since it reflects the real-world observations, but there is no specific limitation in extending our visualization to UKESM or other datasets.

\para{Overall spatial distribution.}
Spatial patterns of blocking event frequency are fundamental to blocking climatology and model evaluation, as emphasized by prior studies~\cite{Nowack2021,Pinheiro2019}.
Motivated by this, we built a frequency heatmap by running our detection method on every day in the analysis period of an ensemble. 
At each grid point, we compute the frequency of days detected as blocked from the normalized $Z_{500}$ anomaly field using the same threshold and domain as in the case studies. 
The resulting frequency heatmap highlights coherent regional organization, offering a compact summary of where blocking events preferentially occur. 

The frequency heatmap showing the spatial distribution of blocking events is presented in~\cref{fig:era5-frequency}. 
Overall, the spatial blocking frequency is distributed according to the latitude. Higher latitude regions, especially Scandinavia and the North Atlantic, have a high frequency of blocking events.
As the latitude decreases, the blocking frequency decreases.
This aligns with the overall observation for the frequency heatmaps from the DG83 in the right of~\cref{fig:era5-frequency}.
However, we notice that the overall frequency from our detection method is higher than that of DG83. This is expected because our method uses a lower scalar threshold than DG83 when detecting blocking events, which typically obtains larger areas for high-pressure systems.

\para{Contour boxplot for daily blocking distribution.}
In addition to the overall summary, we are interested in the variation of the spatial distribution on different days of the year, which may follow different patterns due to seasonal variations.
To better analyze such variation, for each specific day across 41 years in the ERA5 dataset, we compute the contour boxplot and frequency heatmap for blocking events.

\cref{fig:boxplot-daily} presents a series of daily contour boxplots for June 9-20, computed with a number of daily ensembles. For each calendar day, the panel pairs a frequency heatmap with a contour boxplot of detected blocking boundaries, both computed across the 41-year dataset. 
Note that the contributing years can differ from day to day, since not every year exhibits a blocking event on every date.

We observe day-to-day shifts of the median contour in \cref{fig:boxplot-daily}. 
From June 9–11, it remains at higher latitudes, mainly over Scandinavia, the North Atlantic, and the Arctic. 
On June 12, it moves southward, spanning western, central, and parts of eastern Europe. 
During June 13–17, it lingers over western Europe, especially the U.K. and France.
It then shifts eastward on June 18–20, centering over France and Germany.

The accompanying frequency heatmaps show a related evolution in blocking frequency. 
On June 9–13, the frequency pattern is diffuse, with no clear signature of the June 12 southward shift. 
On June 14–17, higher frequencies concentrate over western Europe, consistent with the median contour’s placement.
On June 18–20, the frequency increases over central and eastern Europe, tracking the eastward move of the median contour.

Taken together, these visualizations suggest a tentative mid-June progression in this multi-decadal archive: early-period blocking spread across mid–high latitudes, a mid-month focus over western Europe, and a late-period increase over central and eastern Europe. 
Given the roughly four decades of data and varying year-to-year participation for each date, we interpret these as indicative patterns rather than definitive climatological features.

\para{Spatiotemporal summaries with 3D temporal stacks.}
Beyond daily 2D visualizations (contour boxplots and frequency heatmaps), we summarize blocking behavior using two complementary 3D temporal stacks (\cref{fig:boxplot-stack}). Time is mapped to the vertical axis (May 28-Sep 4). On the left, the 3D median stack displays daily median contours (blocking boundaries) from the contour boxplots stacked along the time axis; the orange strands represent the daily median contours. On the right, the 3D frequency stack combines daily frequency heatmaps into a single 3D volume, where intensity encodes how often blocking events occur at each location over time.

These 3D temporal stacks reveal the broad space-time organization of blocking events. For example, the red box on the left of \cref{fig:boxplot-stack} highlights June 13-20, when the median contours cluster over western Europe, consistent with the daily panels in \cref{fig:boxplot-daily}. Such aggregations provide a compact, qualitative view of when and where blocking events concentrate during the season.

Read across days, the median stack illustrates shifts in the central tendency of blocking footprints rather than the continuous motion of a single event. Vertically coherent segments in the median stack therefore indicate multi-day consistency in the typical blocking location, whereas day-to-day lateral shifts capture seasonal changes in where blocking most frequently occurs. The frequency stack complements this view by revealing bands of elevated occupancy over time: thin, faint layers denote scattered, short-lived events, while thicker, brighter layers mark periods when blocking repeatedly forms in similar regions.

These visualizations also have limitations: projecting a 3D object into 2D can obscure structure. In particular, the 3D frequency stack may hide interior features due to occlusion. In practice, interpretation is most effective with interactive exploration—rotating and slicing through selected dates or latitude–longitude planes—and linking the stacks with the corresponding 2D visualizations.
A video demonstration for exploring the 3D temporal stacks is available in the GitHub repository.

\para{Monthly blocking event frequency.}
Beyond day-to-day variability, a monthly view offers a more robust picture of seasonal behavior for blocking events. Although our parameters are tuned using data from the summer months (the JJA window with ground-truth labels), the agreement with DG83 outside JJA (\cref{tab:agreement}) remains comparable, so we extend the analysis to a monthly basis.

\cref{fig:boxplot-monthly} shows the resulting frequency heatmaps with a monthly ensemble: winter months exhibit substantially higher blocking frequencies than the summer months, with December–January maxima focused over northern Europe and the adjacent North Atlantic, and a more scattered distribution by February–March as circulation transitions toward spring.
This is in line with the expected increase in weather variability compared to December and January, for example due to the weakening of the stratospheric polar vortex and related changes in the frequency of sudden stratospheric warmings (SSWs, major events commonly defined by a reversal of the zonal-mean zonal wind at 60° N and 10 hPa). SSWs are known to be associated with changes in tropospheric westerlies and blocking occurrence~\cite{Baldwin2021}. More generally, blocking is closely associated with the North Atlantic jet: shifts in jet latitude and changes in jet strength are typically accompanied by changes in where blocking occurs and how often it occurs~\cite{Woollings2010,Kautz2022,Wazneh2021}.
In summer, detected blocking events concentrate at higher latitudes (notably north of Europe) with comparatively low frequencies along the Mediterranean coast, reflecting the seasonal shift of blocking events.

%% file: tab-detection-gtd.tex
\begin{table*}[!ht]
\caption{Detection statistics across three detection methods, with respect to the ground-truth labels for both the ERA5 and the UKESM datasets during the JJA period. The best performance for each metric is in bold.
In the ground-truth labels, $33.4\%$ days are blocked for the ERA5 dataset (1979-2019) and $29.0\%$ days are blocked for the UKESM dataset (1960-2060), respectively.}
\vspace{-2mm}
\centering
\resizebox{0.9\textwidth}{!}{
\begin{tabular}{c|ccccc|ccccc}
\hline
\multirow{2}{*}{Methods} & \multicolumn{5}{c|}{ERA5}                                                    & \multicolumn{5}{c}{UKESM}                                                    \\ \cline{2-11} 
       & Days blocked & Accuracy      & Precision     & Recall        & F1-score            & Days blocked & Accuracy      & Precision     & Recall        & F1-score            \\ \hline
DG83   & 32.3\%       & 0.82          & 0.72          & 0.72          & 0.72          & 25.9\%       & 0.87          & 0.80          & 0.72          & 0.75          \\
SOM-BI & 34.6\%       & 0.82          & \textbf{0.73} & 0.75          & 0.74          & 29.6\%       & 0.83          & 0.70          & 0.72          & 0.71          \\
ours   & 37.3\%       & \textbf{0.84} & 0.72          & \textbf{0.83} & \textbf{0.77} & 29.4\%         & \textbf{0.90} & \textbf{0.81} & \textbf{0.83} & \textbf{0.82} \\ \hline
\end{tabular}
}
\vspace{-3mm}
\label{tab:detection}
\end{table*}

\begin{table}[!ht]
    \centering
    \caption{The disagreement in detection results between DG83 and our method for data within the JJA period.}
    \vspace{-2mm}
    \resizebox{0.9\columnwidth}{!}{
\begin{tabular}{c|c|cccc}
\hline
Dataset                & GTD labels   & \begin{tabular}[c]{@{}c@{}}only ours\\ correct\end{tabular} & \begin{tabular}[c]{@{}c@{}}only DG83 \\ correct\end{tabular} & both correct & both incorrect \\ \hline
\multirow{2}{*}{ERA5}  & Blocked     & 161                                                         & 14                                                           & 935          & 213            \\
                       & Not blocked & 100                                                         & 161                                                          & 2204         & 312            \\ \hline
\multirow{2}{*}{UKESM} & Blocked     & 440                                                         & 109                                                          & 1939         & 361            \\
                       & Not blocked & 305                                                         & 340                                                          & 6124         & 280            \\ \hline
\end{tabular}
}
    \label{tab:improvement}
\end{table}

%% file: tab-dg83-alignment.tex
\begin{table}[!ht]
\vspace{-2mm}
\caption{The agreement of our detection results with respect to the DG83 detection results for different months.}
\vspace{-2mm}
\centering
\resizebox{0.9\columnwidth}{!}{
\begin{tabular}{c|ccc|ccc}
\hline
\multirow{2}{*}{Month} & \multicolumn{3}{c|}{ERA5}     & \multicolumn{3}{c}{UKESM}     \\ \cline{2-7} 
                       & Precision & Recall & F1-score & Precision & Recall & F1-score \\ \hline
01                     & 0.741     & 0.968  & 0.840    & 0.675     & 0.936  & 0.784    \\
02                     & 0.746     & 0.970  & 0.843    & 0.686     & 0.943  & 0.794    \\
03                     & 0.804     & 0.963  & 0.876    & 0.690     & 0.927  & 0.791    \\
04                     & 0.800     & 0.931  & 0.860    & 0.705     & 0.954  & 0.811    \\
05                     & 0.773     & 0.932  & 0.845    & 0.693     & 0.893  & 0.780    \\
06                     & 0.823     & 0.958  & 0.885    & 0.695     & 0.828  & 0.756    \\
07                     & 0.807     & 0.911  & 0.855    & 0.779     & 0.849  & 0.812    \\
08                     & 0.740     & 0.886  & 0.806    & 0.754     & 0.856  & 0.802    \\
09                     & 0.797     & 0.896  & 0.844    & 0.700     & 0.845  & 0.766    \\
10                     & 0.777     & 0.959  & 0.858    & 0.587     & 0.897  & 0.710    \\
11                     & 0.669     & 0.925  & 0.777    & 0.660     & 0.956  & 0.781    \\
12                     & 0.778     & 0.977  & 0.866    & 0.689     & 0.948  & 0.798    \\ \hline
\end{tabular}
}
\vspace{-4mm}
\label{tab:agreement}
\end{table}

%% file: sec-discussion.tex
\input{tab-comparison-table}
\section{Conclusion and Discussion}
\label{sec:discussion}

We present an uncertainty visualization framework for detecting and characterizing atmospheric blocking events.
We compare our method against two baseline blocking indices, as summarized in~\cref{tab:compare}.
We introduce a geometry-based detection and tracking method applied to the normalized $Z_{500}$ anomaly field that enforces quasi-stationarity and returns explicit spatial footprints for blocking events. We evaluate the method on both pre-industrial climate simulations (UKESM) and reanalysis data (ERA5). Compared with two strong baselines—DG83 and SOM-BI—our approach achieves higher F1-scores, yielding improved detection performance relative to expert-labeled ground truth. Beyond the tuning season (JJA), monthly comparisons with DG83 show no notable drop in agreement, indicating stable performance across seasons. We further demonstrate results on the well-documented 2003 European heatwave, where the diagnosed blocked days and footprints closely match independent analyses of the event’s evolution.

Another main contribution of our work is a suite of uncertainty-aware summaries for visualizing blocking events. Given an ensemble of detections, we employ contour boxplots to represent typical boundaries and their spread, frequency heatmaps to reveal spatial and temporal occurrence patterns, and 3D temporal stacks to highlight seasonal evolution at a glance. Together, these visualizations make the behavior of blocking events interpretable in both space and time without increasing model complexity.

Our framework has several limitations. The 3D temporal stacks are most effective when explored interactively or through slicing, as static volume renderings may obscure internal structures. In addition, our detection method shows a slight positive bias on the ERA5 dataset, reflected in an imbalance between precision and recall; this bias could be mitigated through further parameter tuning or additional quasi-stationarity constraints while maintaining transparency and reproducibility.

We envision this framework as a valuable tool for climate scientists and meteorologists: by quantifying how blocking frequency, duration, and intensity vary across regions and climate scenarios, it supports more accurate assessment of present and future risks associated with prolonged extreme weather events.

%% file: tab-comparison-table.tex
\begin{table}[!ht]
\centering
\caption{Comparison summary of DG83, SOM-BI, and our method for atmospheric blocking event detection and characterization.}
\vspace{-2mm}
\centering
\begin{minipage}{1.0\columnwidth}
\resizebox{1.0\textwidth}{!}{
\begin{tabular}{c|cccc}
\hline
\textbf{Method} & \textbf{\begin{tabular}[c]{@{}c@{}}Event geometry \&\\ trajectory\end{tabular}} & \textbf{Uncertainty summary}                                                                    & \textbf{\begin{tabular}[c]{@{}c@{}}F1-score \\ (ERA5)\end{tabular}} & \textbf{\begin{tabular}[c]{@{}c@{}}F1-score \\ (UKESM)\end{tabular}} \\ \hline
DG83            & Yes\textsuperscript{*}                                                                                     & Frequency heatmap                                                                               & 0.72                                                                & 0.75                                                                 \\ \hline
SOM-BI          & No                                                                                       & None                                                                                            & 0.74                                                                & 0.71                                                                 \\ \hline
ours            & Yes                                                                                      & \begin{tabular}[c]{@{}c@{}}Contour boxplot; frequency\\ heatmap; 3D temporal stack\end{tabular} & \textbf{0.77}                                                       & \textbf{0.82}                                                        \\ \hline
\end{tabular}
} \\
\footnotesize{\textsuperscript{*} DG83 requires additional post-processing for objects detection and tracking.}
\end{minipage}
\vspace{-6mm}
\label{tab:compare}
\end{table}

%% file: sec-runtime.tex
\section{Runtime Performance and Analysis}
\label{sec:runtime}

In this section, we report the runtime performance of the blocking-detection and uncertainty-visualization components of our framework. The benchmarking hardware configuration is described in \cref{sec:implementation}.

\para{Blocking detection.}
We report the runtime of blocking detection on both ERA5 and UKESM datasets in~\cref{tab:runtime}. 
For comparison, we include the runtime of the DG83 pipeline with the modifications of Pinheiro et al.~\cite{DG83,Pinheiro2019}; see \cref{sec:baseline} for details.

\cref{tab:runtime} shows that both DG83 and our method require less than one minute to detect blocking events for the JJA days across both datasets.
Both methods are highly efficient, as they do not involve a training process to learn blocking patterns.
Compared with DG83, our implementation incurs an additional runtime of $43.1\%$–$46.7\%$, primarily due to the extra step of searching for large region overlaps.
 
\input{tab-runtime}

\para{Uncertainty visualization.}
Recall that we visualize uncertainty using blocking frequency heatmaps, contour boxplots, and 3D temporal stacks. Computing frequency maps and temporal stacks involves straightforward accumulation, whereas contour boxplots are more computationally demanding: for a given calendar date~$i$ with~$d_i$ blocked days across all years, constructing the contour boxplot requires~$O(d_i^2)$ operations.

For all data within the JJA period, our implementation takes 35.18~seconds on ERA5 (41~years) and 140.31~seconds on UKESM (101~years) for all uncertainty-visualization computations. The higher cost for UKESM is expected, as the typical~$d_i$ is larger than in ERA5. In both datasets, the runtime remains sufficiently low for practical use and downstream analysis.

\para{Potential improvement.}~Although our implementation is already efficient, further performance gains can be achieved through parallelization. In blocking detection, both region-overlap tests and trajectory linkage are independent across time steps and thus well suited to parallel execution. For uncertainty visualization, particularly contour boxplots, the pairwise band computations are also parallelizable. These opportunities make it straightforward to scale our framework to larger datasets.

%% file: tab-runtime.tex
\begin{table}[!ht]
\vspace{-2mm}
\caption{Runtime (in seconds) for blocking detection during the JJA period using DG83 and our method.}
\vspace{-2mm}
\resizebox{0.95\columnwidth}{!}{
\begin{tabular}{c|c|c|c|c}
\hline
\textbf{Dataset}       & \textbf{Years}       & \textbf{Method} & \textbf{\begin{tabular}[c]{@{}c@{}}Total runtime \\ (sec)\end{tabular}} & \textbf{\begin{tabular}[c]{@{}c@{}}Average runtime per year \\ (sec)\end{tabular}} \\ \hline
\multirow{2}{*}{ERA5}  & \multirow{2}{*}{41}  & DG83            & 31.02                                                                   & 0.757                                                                              \\
                       &                      & Ours            & 45.51                                                                   & 1.110                                                                              \\ \hline
\multirow{2}{*}{UKESM} & \multirow{2}{*}{101} & DG83            & 18.17                                                                   & 0.180                                                                              \\
                       &                      & Ours            & 26.00                                                                   & 0.257                                                                              \\ \hline
\end{tabular}
}
\vspace{-2mm}
\label{tab:runtime}
\end{table}

%% file: sec-extended-eval.tex
\section{Extended Evaluation}
\label{sec:extended-eval}

In this section, we extend the blocking-detection evaluation to assess temporal stability across the JJA season. As in~\cref{sec:detection}, we evaluate performance against the expert GTD labels. We omit SOM-BI here due to the unavailability of per-day outputs in our environment.

\begin{figure}[!ht]
    \centering
    \includegraphics[width=0.98\columnwidth]{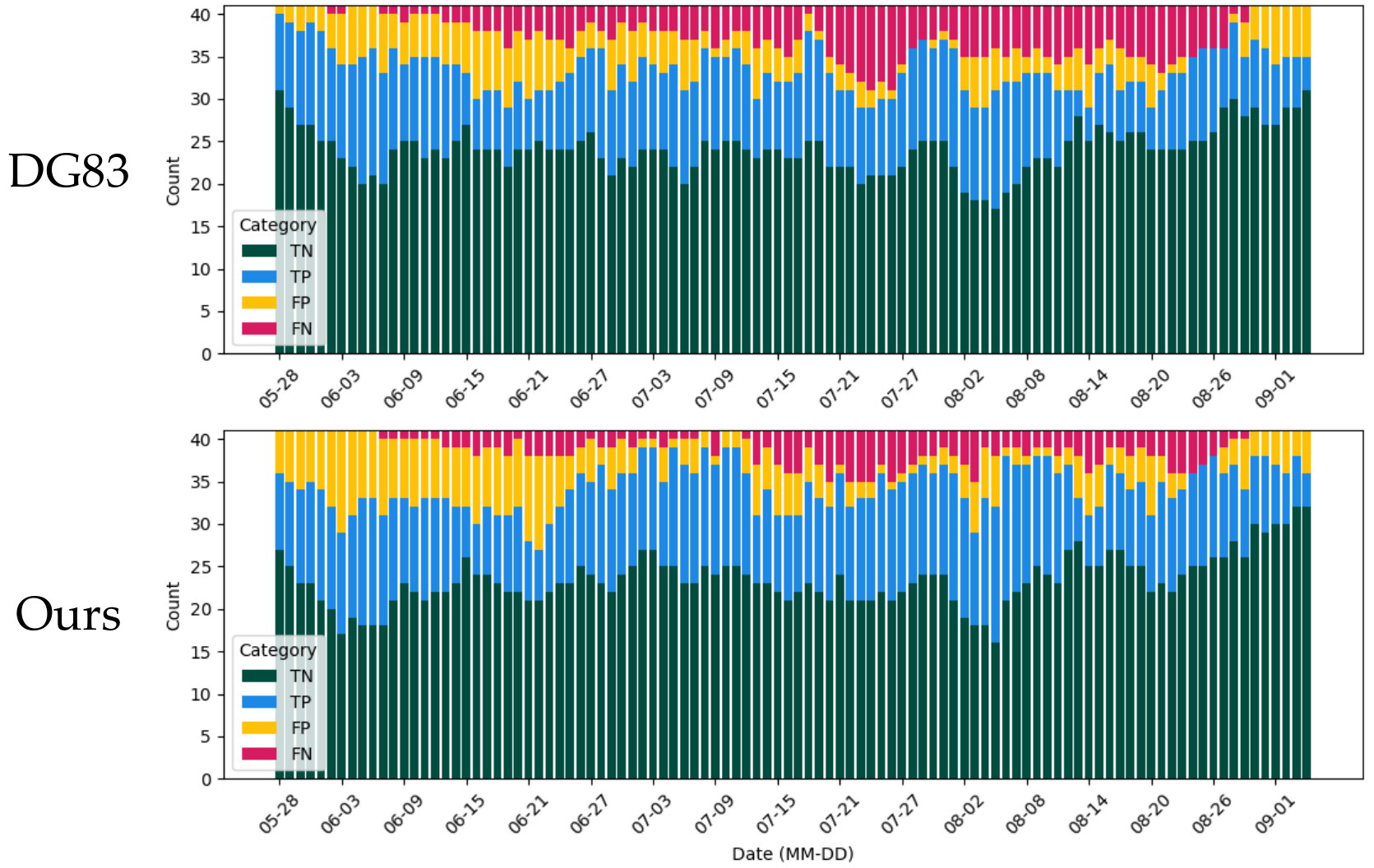}
    \vspace{-3mm}
    \caption{Temporal breakdown of daily detection outcomes during the JJA period in ERA5 (41 years).
Top: DG83; bottom: ours. For each calendar date (x-axis), the stacked bar aggregates the 41 daily instances (one per year), so all bars have equal height. Colors indicate counts of TN (dark green), TP (blue), FP (yellow), and FN (magenta). This visualization compares how the composition of correct and error outcomes evolves through the season for the two methods.} 
    \label{fig:temporal-eval-ERA5}
    \vspace{-3mm}
\end{figure}

\para{Seasonal pattern of daily outcomes.}
\cref{fig:temporal-eval-ERA5} aggregates, for each calendar date, the 41 JJA instances from ERA5; all bars have equal height, and the color composition indicates the proportions of \textbf{TN}, \textbf{TP}, \textbf{FP}, and \textbf{FN} (true negatives, true positives, false positives, and false negatives, respectively).

In early summer (May~28–June~24), our method (bottom) shows a larger share of \textbf{FP} than \textbf{FN}, reflecting a more inclusive tendency during this period. From mid- to late-summer (June~25–September~4), \textbf{FP} decrease noticeably, while a modest but persistent number of \textbf{FN} appear from mid-July through late-August. Because \textbf{FP} and \textbf{FN} remain of similar magnitude during this interval, the behavior is broadly balanced rather than biased toward misses or false alarms. In contrast, DG83 (top) records fewer \textbf{FP} early in the season but substantially more \textbf{FN} overall, especially through mid- and late-summer, consistent with a more conservative detector. Both methods exhibit a seasonal pattern within JJA: \textbf{FP} generally decline from early to late summer, while \textbf{FN} increase. These represent multi-week patterns observable at the 41-year aggregate level; we do not interpret day-scale fluctuations.

\begin{figure}[!ht]
    \centering
    \vspace{-3mm}
    \includegraphics[width=0.98\columnwidth]{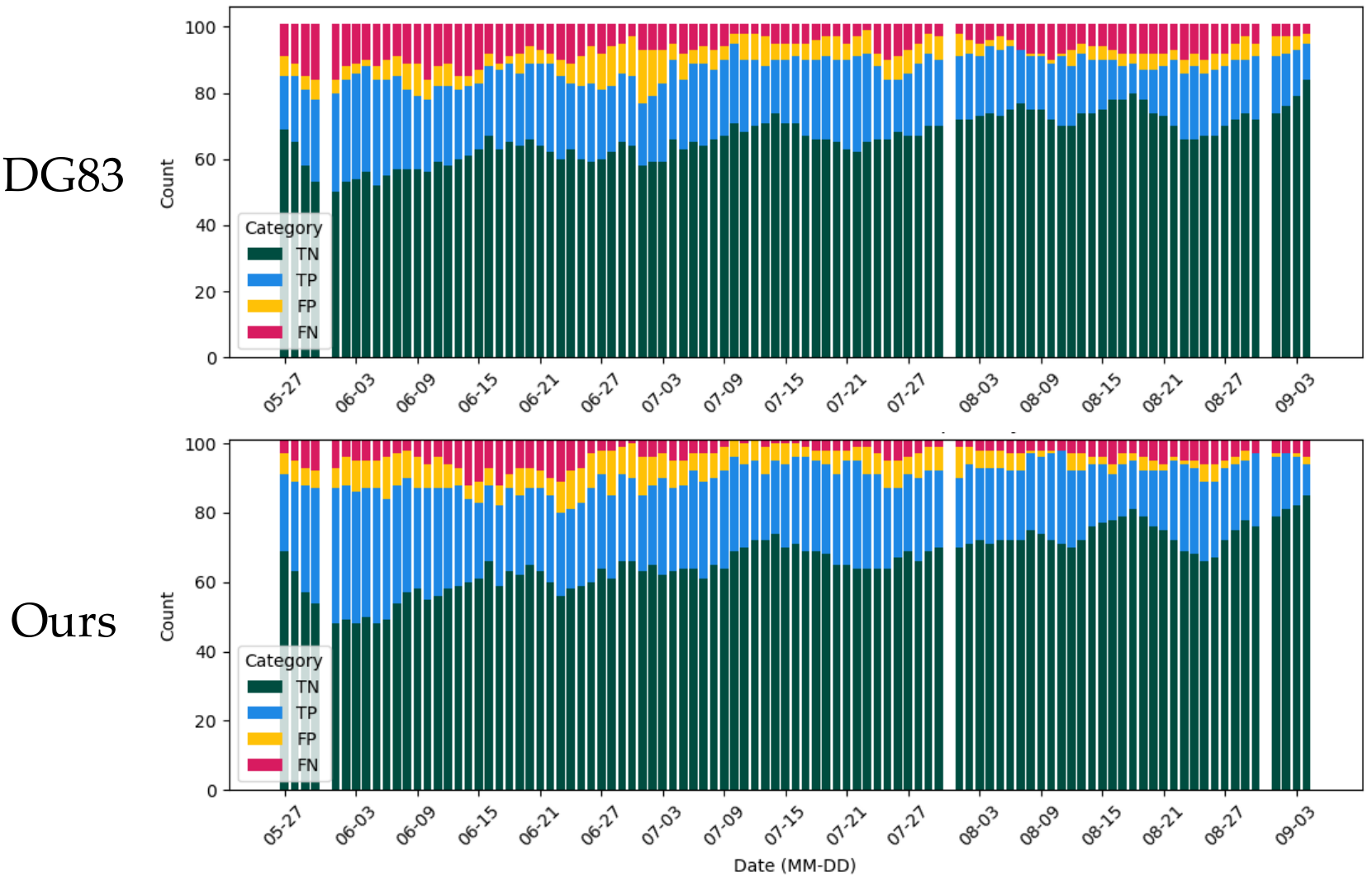}
    \vspace{-3mm}
    \caption{Temporal breakdown of daily detection outcomes during the JJA period in UKESM (101 years).
Top: DG83; bottom: ours. For each calendar date (x-axis), the stacked bar aggregates the 101 daily instances (one per year), so all bars have equal height. Colors indicate counts of TN (dark green), TP (blue), FP (yellow), and FN (magenta).
Columns for May 31, July 31, and August 31 are blank because the UKESM simulation calendar does not include these dates.
    }
    \vspace{-4mm}
    \label{fig:temporal-eval-UKESM}
\end{figure}

\cref{fig:temporal-eval-UKESM} presents the corresponding per-date aggregation for UKESM (101 JJA seasons). Unlike ERA5, the TN/TP/FP/FN composition remains comparatively uniform throughout the season, showing little systematic intra-seasonal drift—consistent with the pre-industrial simulation setting, which lacks pronounced seasonal variation. Our method (bottom) maintains a balanced mix of \textbf{FP} and \textbf{FN} across dates, indicating no strong bias toward false alarms or misses. Compared with DG83 (top), our detector produces a similar number of \textbf{FP} but noticeably fewer \textbf{FN}, consistent with its higher overall detection performance on UKESM.

Using a single parameter configuration, our detector exhibits stable performance across both JJA seasons and datasets, while accommodating mild seasonal variation. In ERA5, early summer shows a higher proportion of \textbf{FP}, an inclusive phase consistent with our previously noted bias toward higher recall, followed by mid- to late-summer, where \textbf{FP} and \textbf{FN} remain broadly balanced. In UKESM, the per-date composition is nearly flat, maintaining a consistent balance between \textbf{FP} and \textbf{FN} throughout the season. Together, these results indicate that our detector operates stably at the seasonal scale, slightly favoring inclusivity in early JJA and achieving balanced performance thereafter, while maintaining the precision levels reported in the main evaluation.

